\documentclass[]{aastex63}
\usepackage{multirow}
\usepackage{graphicx}
\usepackage{amsmath,amssymb}
\usepackage{longtable}
\graphicspath{{./}{figures/}}

\begin{document}

\title{Very-High-Energy Gamma-Ray Afterglows of GRB~201015A and GRB~201216C}
\correspondingauthor{En-Wei Liang}
\author[0000-0003-0726-7579]{Lu-Lu Zhang}
\affil{Guangxi Key Laboratory for Relativistic Astrophysics, School of Physical Science and Technology, Guangxi University, Nanning 530004, China}
\author[0000-0002-9037-8642]{Jia Ren}
\affil{School of Astronomy and Space Science, Nanjing University, Nanjing 210023, China}
\affil{Key Laboratory of Modern Astronomy and Astrophysics (Nanjing University), Ministry of Education, Nanjing 210023, China}
\author[0000-0002-8385-7848]{Yun Wang}
\affiliation{Key Laboratory of Dark Matter and Space Astronomy, Purple Mountain Observatory, Chinese Academy of Sciences, Nanjing 210034, China}
\affiliation{School of Astronomy and Space Science, University of Science and Technology of China, Hefei, Anhui 230026, China}
\author[0000-0002-7044-733X]{En-Wei Liang*}
\affil{Guangxi Key Laboratory for Relativistic Astrophysics, School of Physical Science and Technology, Guangxi University, Nanning 530004, China}
\begin{abstract}
Gamma-ray bursts (GRBs) 201015A and 201216C are valuable cases with detection of very high energy (VHE) gamma-ray afterglows. By analysing their prompt emission data, we find that GRB 201216C is an extremely energetic long GRB with a hard gamma-ray spectrum, while GRB 201015A is a relative sub-energetic, soft spectrum GRB. Attributing their radio-optical-X-ray afterglows to the synchrotron radiation of the relativistic electrons accelerated in their jets, we fit their afterglow lightcurves with the standard external shock model and infer their VHE afterglows from the synchrotron self-Compton scattering process of the electrons.  It is found that the jet of GRB 201015A is mid-relativistic ($\Gamma_0=44$) surrounded by a very dense medium ($n=1202$~cm$^{-3}$) and the jet of GRB~201216C is ultra-relativistic ($\Gamma_0=331$) surrounded by a moderate dense medium ($n=5$~cm$^{-3}$). The inferred peak luminosity of the VHE gamma-ray afterglows of GRB 201216C is approximately $10^{-9}$~erg~cm$^{-2}$~s$^{-1}$ at $57-600$ seconds after the GRB trigger, making it can be detectable with the MAGIC telescopes at a high confidence level, even the GRB is at a redshift of 1.1. Comparing their intrinsic VHE gamma-ray lightcurves and spectral energy distributions with GRBs~180720B, 190114C, and 190829A, we show that their intrinsic peak luminosity of VHE gamma-ray afterglows at $10^{4}$~seconds post the GRB trigger is variable from $10^{45}$ to $5\times 10^{48}$~erg~s$^{-1}$, and their kinetic energy, initial Lorentz factor, and medium density are diverse among bursts.
\end{abstract}


\section{Introduction}
\label{sec:introduction}
The very high energy (VHE, $E>0.1$ TeV) afterglows of gamma-ray bursts (GRBs) were previously theoretically predicted (e.g., \citealp{Meszaros_1993_Rees_ApJ...405..278M, Waxman_1997_ApJ...485L...5W, Dermer_2000_Chiang_ApJ...537..785D, Sari_2001_Esin_ApJ...548..787S, Zhang_2001_Meszaros_ApJ...559..110Z, Wang_2001_Lu_ApJ...556.1010W, Meszaros_2004_Razzaque_NewAR..48..445M, Fan_2008_Piran_FrPhC...3..306F}) and recently detected with the ground-based telescopes for several GRBs (e.g., \citealp{Abdalla-2019-Adam-Natur.575..464A, MAGIC_Collaboration_2019_Acciari_Natur.575..455M, MAGIC_Collaboration_2019_Acciari_Natur.575..459M, Acciari-2021-Ansoldi-ApJ...908...90A, H._E._S._S._Collaboration_2021_Abdalla_Sci...372.1081H}).   
The VHE gamma-ray afterglows of GRB 180720B were firstly detected with the High Energy Stereoscopic System (H.E.S.S.) in a confidence level of $5.3\sigma$ in the $0.1-0.4$~TeV band \citep{Vreeswijk_2018_Kann_GCN.22996....1V, Abdalla-2019-Adam-Natur.575..464A}. GRB~180720B is an extremely bright GRB, which has an isotropic gamma-ray energy of $E_{\rm\gamma,iso}=6\times 10^{53}$~erg at redshift $z=0.654$. More excitingly, the VHE gamma-ray afterglows of GRB~190114C were convincingly detected with the Major Atmospheric Gamma Imaging Cerenkov (MAGIC) telescopes in the $0.3-1~{\rm TeV}$ band in a high confidence level of $> 50\sigma$ \citep{MAGIC_Collaboration_2019_Acciari_Natur.575..455M}. GRB~190114C is also a very bright GRB. Its$E_{\rm\gamma,iso}$ is $3\times10^{53}$~erg at redshift $z=0.4245$. Being due to strong extra-galactic background light (EBL) absorption to the sub-TeV/TeV photons from high redshift sources (e.g., \citealp{Razzaque_2009_Dermer_ApJ...697..483R}), VHE gamma-ray afterglows should be preferably detected for low redshift GRBs. Unsurprisingly, the TeV afterglows is firstly detected in nearby GRB~190829A ($z=0.0785$) with H.E.S.S. in the $0.2-4$~TeV band with a confidence level of $21.7\sigma$ \citep{H._E._S._S._Collaboration_2021_Abdalla_Sci...372.1081H}. Different from GRBs~180720B and 190114C, GRB~190829A is a sub-energy, low-luminosity (LL) GRB, which has $L_{\rm iso}= 1.9\times 10^{49}~\rm erg~\rm s^{-1}$ and $E_{\rm\gamma,iso}=2\times 10^{50}$~erg (\citealp{H._E._S._S._Collaboration_2021_Abdalla_Sci...372.1081H, Zhang-2021-Ren-ApJ...917...95Z}), indicating that the afterglow jets of both energetic and sub-energetic GRBs can accelerate particles to an extremely high energy and produce the VHE gamma-ray afterglows (e.g. \citealp{Hurley_1994_Dingus_Natur.372..652H, Murase_2006_Ioka_ApJ...651L...5M}).

It is generally believed that the VHE afterglows of GRBs are attributed to the synchrotron, synchrotron self-Compton (SSC), and/or external inverse-Compton (EIC) radiations of the electrons accelerated in the jets (e.g., \citealp{Dermer_2000_Chiang_ApJ...537..785D, Zhang_2001_Meszaros_ApJ...559..110Z, Sari_2001_Esin_ApJ...548..787S, MAGIC_Collaboration_2019_Acciari_Natur.575..459M, Acciari-2021-Ansoldi-ApJ...908...90A, H._E._S._S._Collaboration_2021_Abdalla_Sci...372.1081H, Zhang-2021-Ren-ApJ...917...95Z}).
The broad-band spectral energy distribution (SED) in the optical, X-ray,
and sub-TeV gamma-ray bands of GRB~190114C can be roughly modeled with the synchrotron
radiations and SSC process of the ultra-relativistic electrons in the jet (e.g., \citealp{Wang_2019_Liu_ApJ...884..117W,Huang_2020_Liang_ApJ...903L..26H,Zhang_2020_Christie_mnras_v496.p974..986, Joshi_2021_Razzaque_MNRAS.505.1718J, Yamasaki_2022_Piran_MNRAS.512.2142Y}).
However, there is controversy over the radiation mechanism of the VHE gamma-ray emission of GRB~190829A. As reported by \cite{H._E._S._S._Collaboration_2021_Abdalla_Sci...372.1081H}, the power-law spectral index of the gamma-rays between 0.18 and 3.3 TeV band is very hard ($\Gamma_\gamma=2.07\pm0.09$). Although the SSC process can explain the VHE gamma-ray flux of GRB~190829A at $t=2\times 10^4$~s after GBM trigger (e.g., \citealp{Zhang_2021_Murase_ApJ...920...55Z}), the hard VHE gamma-ray spectrum challenges this scenario \citep{H._E._S._S._Collaboration_2021_Abdalla_Sci...372.1081H,Huang_2022_Kirk_ApJ...925..182H}.
It was proposed that hadronic processes may play a role in explaining the hard spectrum component \citep{Sahu_2022_Valadez_Polanco_ApJ...929...70S}.

More recently, it was reported that the VHE afterglows of GRB~201015A and GRB~201216C were detected with the MAGIC telescopes \citep{Blanch-2020-Gaug-GCN.28659....1B, Blanch-2020-Longo-GCN.29075....1B}, adding two valuable examples for studying the radiation physics of GRB afterglow jets.
In this paper, we analyze their prompt and afterglow data and model their radio-optical-X-ray afterglows for investigating their jet properties. Furthermore, we compare the jet properties among GRBs~180720B, 190114C, 190829A, 201015A, and 201216C for exploring possible general features of the GRB jets with detection of VHE gamma-ray afterglows.
Our data analysis for GRB~201015A and GRB~201216C is reported in \S~{\ref{data}}.
Modeling of their radio-optical-X-ray afterglows is shown in \S~{\ref{modeling}}. Discussion of current GRBs with detection of VHE afterglows is presented in \S~{\ref{VHE ssc}}.
We summarize our results in \S~{\ref{summary}}.
Throughout, the convention $Q=10^{n}Q_{n}$ in the cgs units are adopted. We take the cosmology
parameters as $H_0=67.8~\rm km~s^{-1}~Mpc^{-1}$, and $\Omega_M=0.308$ \citep{PlanckCollaboration_2016_Ade_aap_v594.p13..13A}.

\section{Multi-wavelength Observations of GRBs 201015A and 201216C}\label{data}
\subsection{GRB 201015A}
\begin{figure}[htbp]
\centering
\includegraphics[angle=0,width=0.45\textwidth]{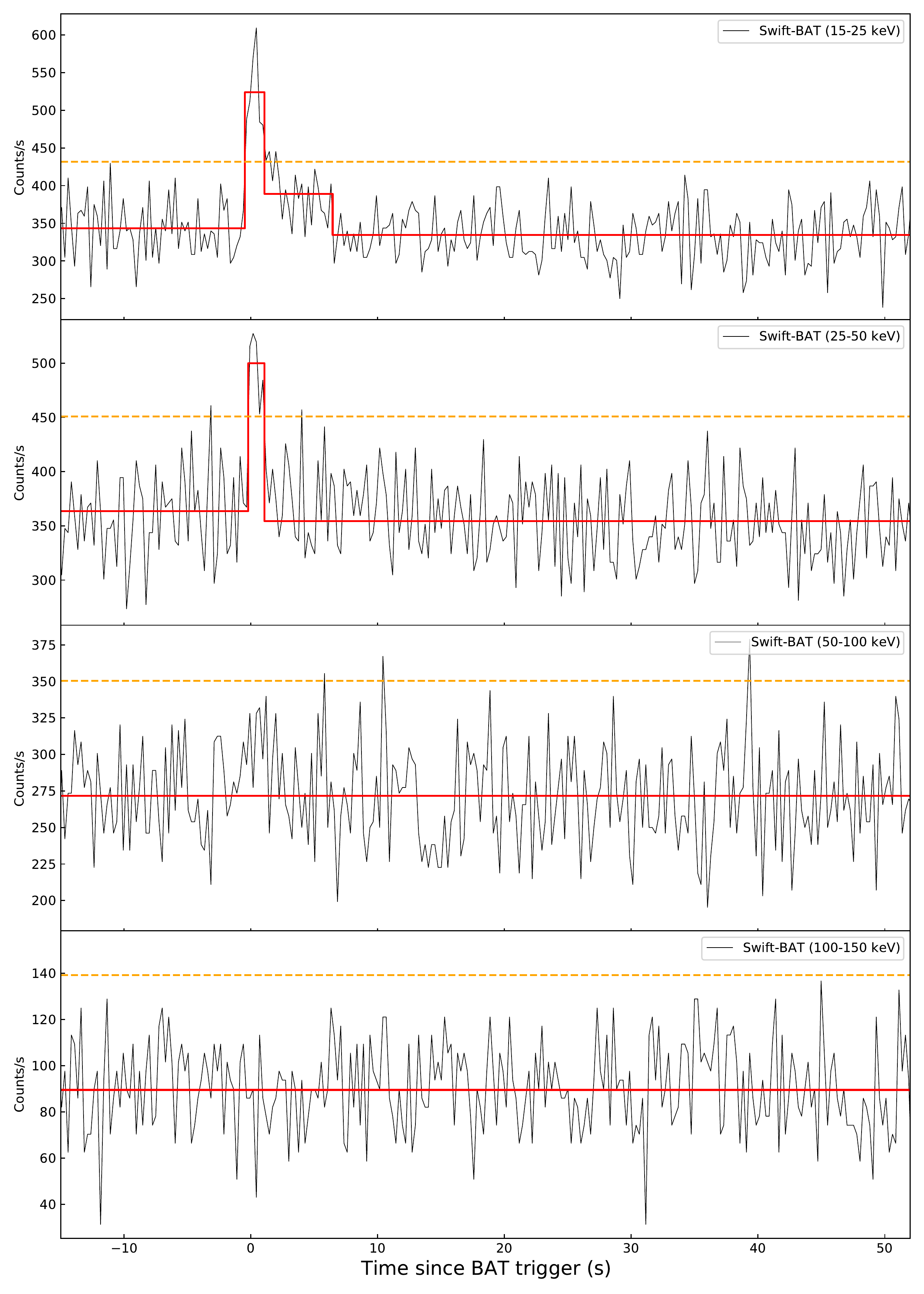}
\includegraphics[angle=0,width=0.45\textwidth]{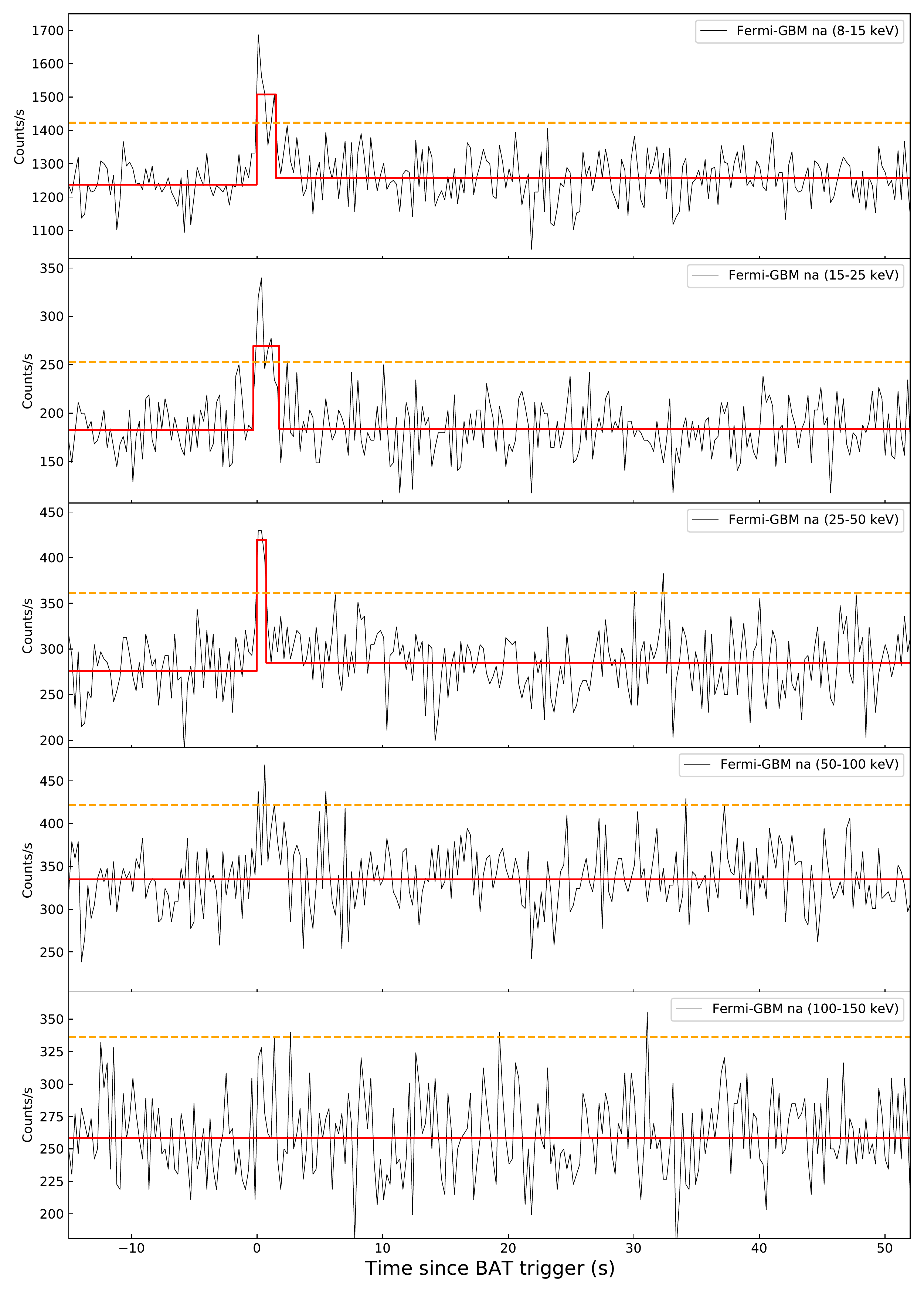}
\caption{Prompt gamma-ray lightcurves of GRB~201015A. The solid red step lines mark the analysis results with the Bayesian block algorithm. The dash orange horizontal lines represents $\rm SNR=5$.}
\label{prompt1}
\end{figure}

GRB~201015A triggered the Burst Alert Telescope (BAT) on board the Swift mission \citep{Markwardt-2020-Barthelmy-GCN.28658....1M}. The spectroscopic observations with the Nordic Optical Telescope show a few absorption features due to
\ion{Mg}{2}, \ion{Mg}{1} and \ion{Ca}{2} multiplets, and reveal its redshift as $z=0.423$ \citep{Izzo-2020-Malesani-GCN.28661....1I}. We download the BAT data from the Swift website\footnote{\url{https://www.swift.ac.uk/archive/selectseq.php?tid=1000452\&source=obs&name=GRB\%20201015A\&reproc=1\&referer=portal}}
and process the data utilizing the BAT official software package.
The left panel of Figure~\ref{prompt1} shows its lightcurves
in the energy bands of $15-25$~keV, $25-50$~keV, $50-100$~keV, and $100-150$~keV.
Our temporal analysis to the lightcurves with the Bayesian Block algorithm \citep{Scargle-2013-Norris-ApJ...764..167S} is also in Figure~\ref{prompt1}.
One can observe that the burst is soft and short.
It is only detected in the $15-25$ and $25-50$~keV bands with a Signal-to-Noise Ratio (SNR) of 5.
It has a sharp pulse with a duration of $\sim 1$~second, followed by a soft extended emission,
which has a duration of $\sim 6$~s in the $15-25$~keV band.
Its duration of the GRB ($T_{90}$) is $\sim 10$~s in the $15-350$~keV band \citep{Markwardt-2020-Barthelmy-GCN.28658....1M}.
Its BAT spectrum is well fitted with a single power-law function,
yielding a photon index of $\Gamma_\gamma=3.03^{+0.78}_{-0.58}$,
indicating the burst is soft.\footnote{The spectrum is also can be fitted with a black body spectrum,
which yields a temperature of $kT=4.79 ^{+1.40}_{-1.06}$~keV.
The $\chi^2_r$/dof of the fit is $46/57$.} 
The reduced $\chi^2$/dof (degree of freedom) of the fit is $44/59$.

GRB~201015A was also sub-threshold detected with Fermi/Gamma-Ray-Monitor (GBM) \citep{Fletcher-2020-Veres-GCN.28663....1F}.
We download the GBM daily data\footnote{\url{https://heasarc.gsfc.nasa.gov/FTP/fermi/data/gbm/daily/2020/10/15/current/}}
and extract its lightcurves in the energy bands of $8-15$~keV, $15-25$~keV, $25-50$~keV, $50-100$~keV, and $100-150$~keV by utilizing the official GBM software package.
We show the lightcurves in the right panel of Figure~\ref{prompt1} in comparison with that observed by BAT.
It is found that the GRB is detected in the $8-15$~keV, $15-25$~keV, and $25-50$~keV bands.
\cite{Fletcher-2020-Veres-GCN.28663....1F} reported that the spectrum of GRB~201015A observed with GBM in the time slice ($\rm{T_{0}}-0.256$~s, $\rm{T_{0}}+0.896$~s) can be fitted with the band function,
yielding a peak energy of the $\nu f_\nu$ spectrum as $E_{\rm peak}= 14\pm 6~\rm keV$ and a high energy photon index as $\beta = -(2.4\pm 0.21)$ by fixing the low energy photon index as $\alpha = -1.0$.
The derived isotropic energy of the prompt emission is
$E_{\gamma,\rm{iso}}=(1.1\pm 0.02)\times 10^{50}~\rm erg$ 
by using these spectral parameters \citep{Minaev-2020-Pozanenko-GCN.28668....1M}.
It bridges the sub-energetic/low-luminosity and energetic/high-luminosity GRBs (e.g., \citealp{Liang_2007_Zhang_ApJ...662.1111L}).

\begin{figure}[htbp]
\centering
\includegraphics[angle=0,width=0.45\textwidth]{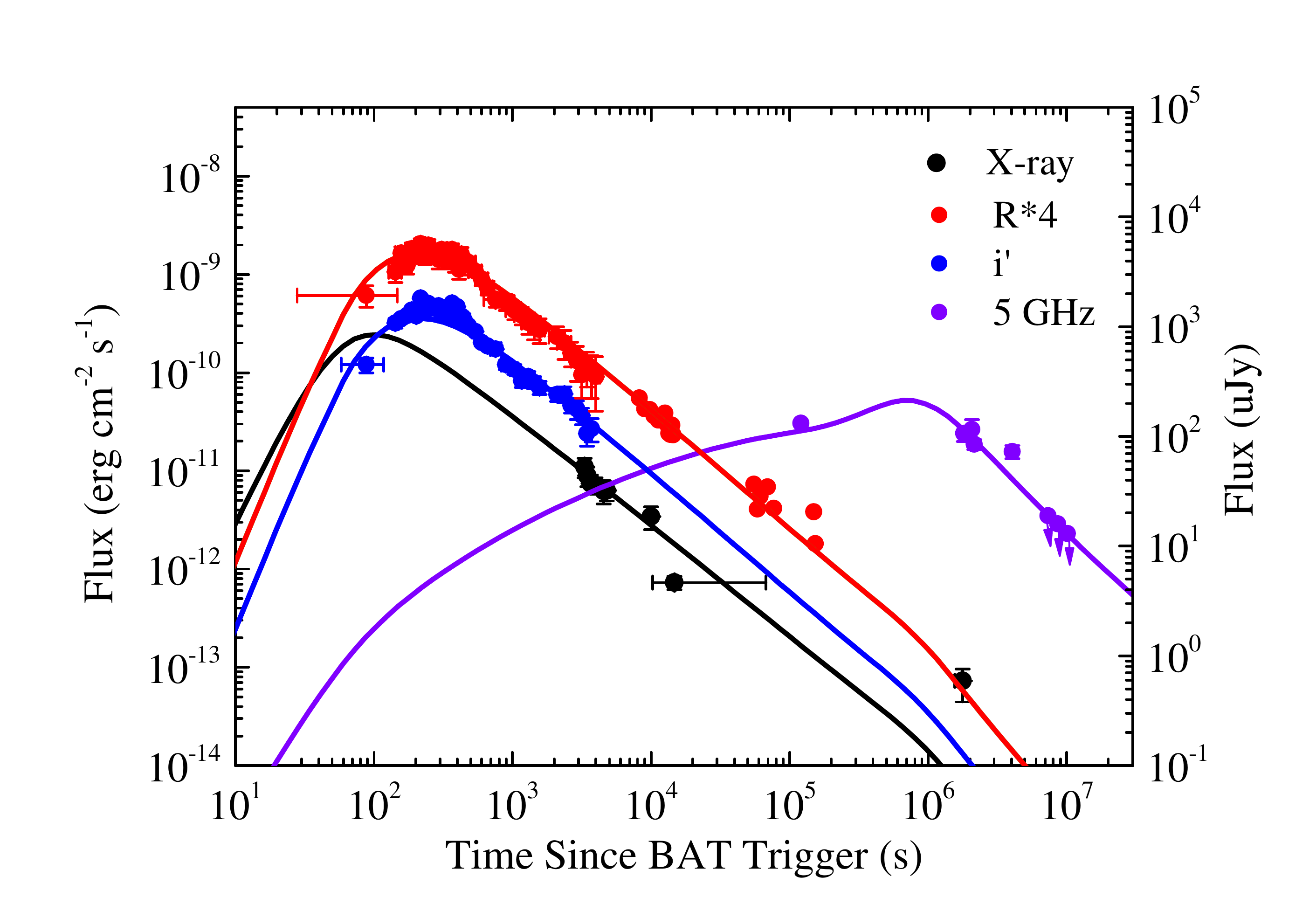}
\includegraphics[angle=0,width=0.45\textwidth]{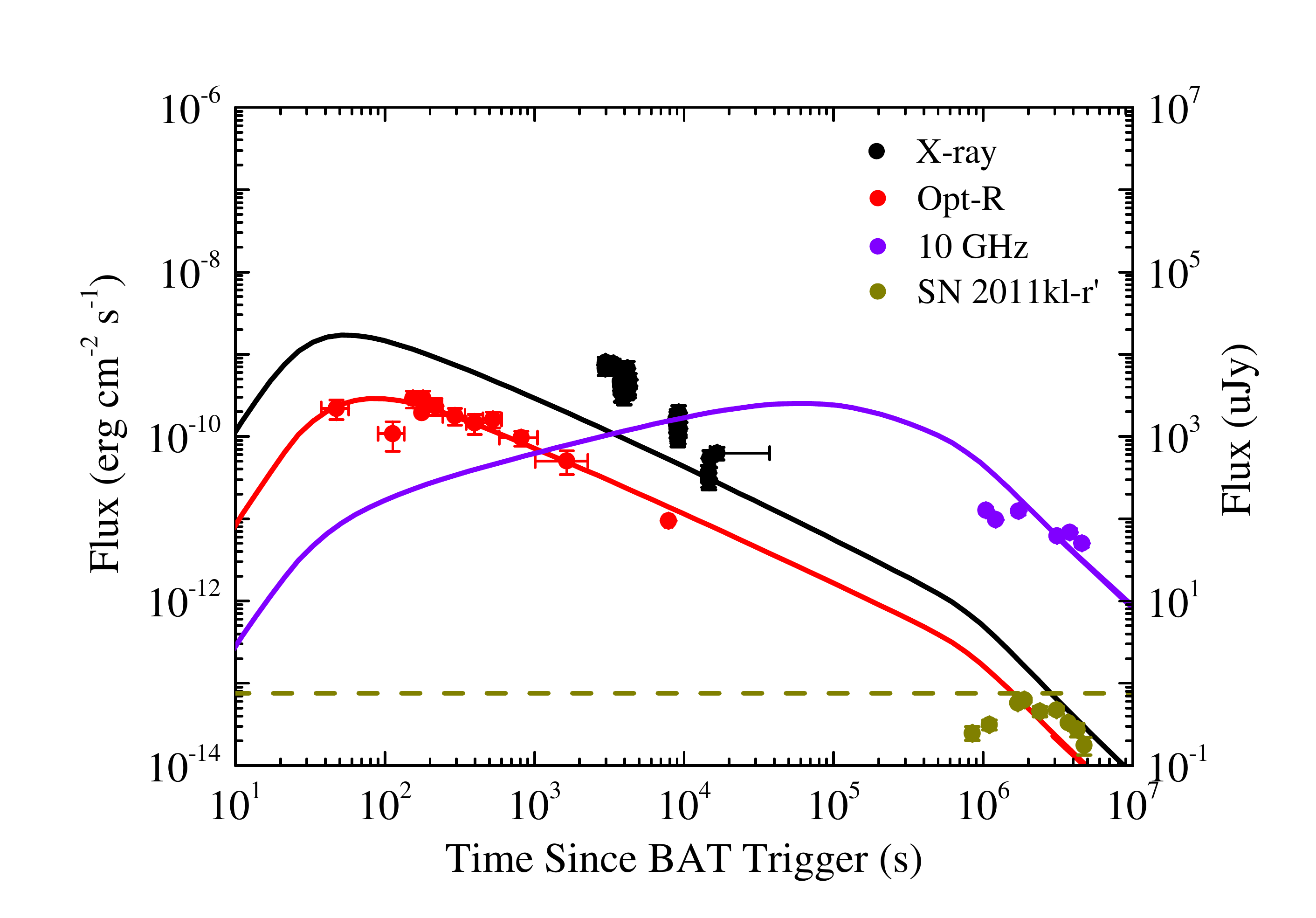}
\caption{
The X-ray, optical, and radio afterglows of GRB~201015A ({\em left panel)} and GRB~201216C ({\em right panel)} together with the fits by the external FS model. In the left-hand panel, the points of the purple down arrows indicate the limits of the radio afterglows.
In the right-hand panel, the points of dark yellow represents the SN 2011kl associated with GRB 111209A in the r' band at $z=1.1$. The dash line dark yellow represents the $24.2~\rm mag$ limit of GROND telescope in an exposure time of 8 minutes.}
\label{201015A}
\end{figure}

The multi-wavelength afterglow lightcurves of GRB~201015A are shown in the left panel of Figure~\ref{201015A}.
Due to an observing constraint,
its X-ray afterglows were detected with the X-Ray Telescope (XRT) on board the Swift mission since $3214.1$~s after the BAT trigger \citep{D'Ai-2020-Gropp-GCN.28660....1D}.
We download the Swift/XRT lightcurve data from the website of Swift Burst Analyzer\footnote{\url{https://www.swift.ac.uk/burst_analyser/01000452/}}.
The X-ray decays as a power-law function with a slope of $\alpha_{\rm X}=1.49_{-0.21}^{+0.24}$ \citep{D'Ai-2020-Gropp-GCN.28660....1D}.
The X-ray afterglow was also detected by the Chandra X-ray telescope at late epochs $8.4$ and $13.6$~days after trigger \citep{Gompertz-2020-Levan-GCN.28822....1G}.
The X-ray flux observed with the Chandra telescope X-ray afterglow is around two orders of magnitude brighter than the extrapolation of the power-law function.
Since the X-ray flux faded between the two Chandra epochs with a power-law index of approximately $-0.8$,
the X-rays should be the afterglows of the GRBs.
The X-ray afterglow is also seen in the Target of Opportunity (ToO) observation with the Swift/XRT in the epoch of $18-21$~days after the BAT trigger \citep{D'Elia-2020-Swift_Team-GCN.28857....1D}.
Similar to the Chandra X-ray telescope observations,
the X-ray flux is also 100 times brighter than the extrapolation of the power-law function.
The shallower decay slope than that derived from data between $0.03$ and $0.8$~days after burst may indicate extra energy injection at the late epochs.
In this analysis, we do not consider any energy injection at late epochs since we are only interested in its early afterglows.

We collect the optical $R$ and $i^{\prime}$ band from the \cite{Pozanenko_2020_Belkin_GCN.29033....1P} and \cite{Komesh-2022-Grossan-arXiv221103029K}, respectively.
Early optical afterglows are monitored in the $g^{\prime}$, $r^{\prime}$, $i^{\prime}$~bands since $58$~seconds after the BAT trigger by
the Nazarbayev University Transient Telescope at Assy-Turgen Astrophysical Observatory \citep{Grossan-2020-Maksut-GCN.28674....1G}.
We make the Galactic extinction correction as $g^{\prime}_{\lambda}=1.119$, $i^{\prime}_{\lambda}=0.575$, and $R_{\lambda}=0.735$ \citep{Schlafly-2011-Finkbeiner-ApJ...737..103S}.
We convert $r^{\prime}$ to the $R$ band by using the expression
$m_{R}=m_{r^{\prime}}-2.5\beta_{O} {\rm log_{10}}(\lambda_{R}/\lambda_{r^{\prime}})+2.5 {\rm log_{10}}(f_{0,R}/f_{0,r^{\prime}})$,
where $f_{0,R}$, $f_{0,r^\prime}$ are the zero fluxes of the $R$ and $r^\prime$ band \citep{Gao-2015-Wang-ApJ...810..160G}, and the optical spectral index $\beta_{O}$ is set as $0.5$.
The optical lightcurves of GRB~201015A show a clear onset bump, peaking at $\sim 200$~s post the GRB trigger.
The early optical and X-ray afterglows have same decay slope post the peak time.
The radio afterglows of GRB~201015A were observed with
the e-MERLIN, VLA and EVN Telescopes at 1.5 and 5~GHz.
We adopt 5~GHz data and some upper limits for our analysis \citep{Giarratana-2022-Rhodes-arXiv220512750G}.

\subsection{GRB 201216C}
\begin{figure}[htbp]
\centering
\includegraphics[angle=0,width=0.45\textwidth]{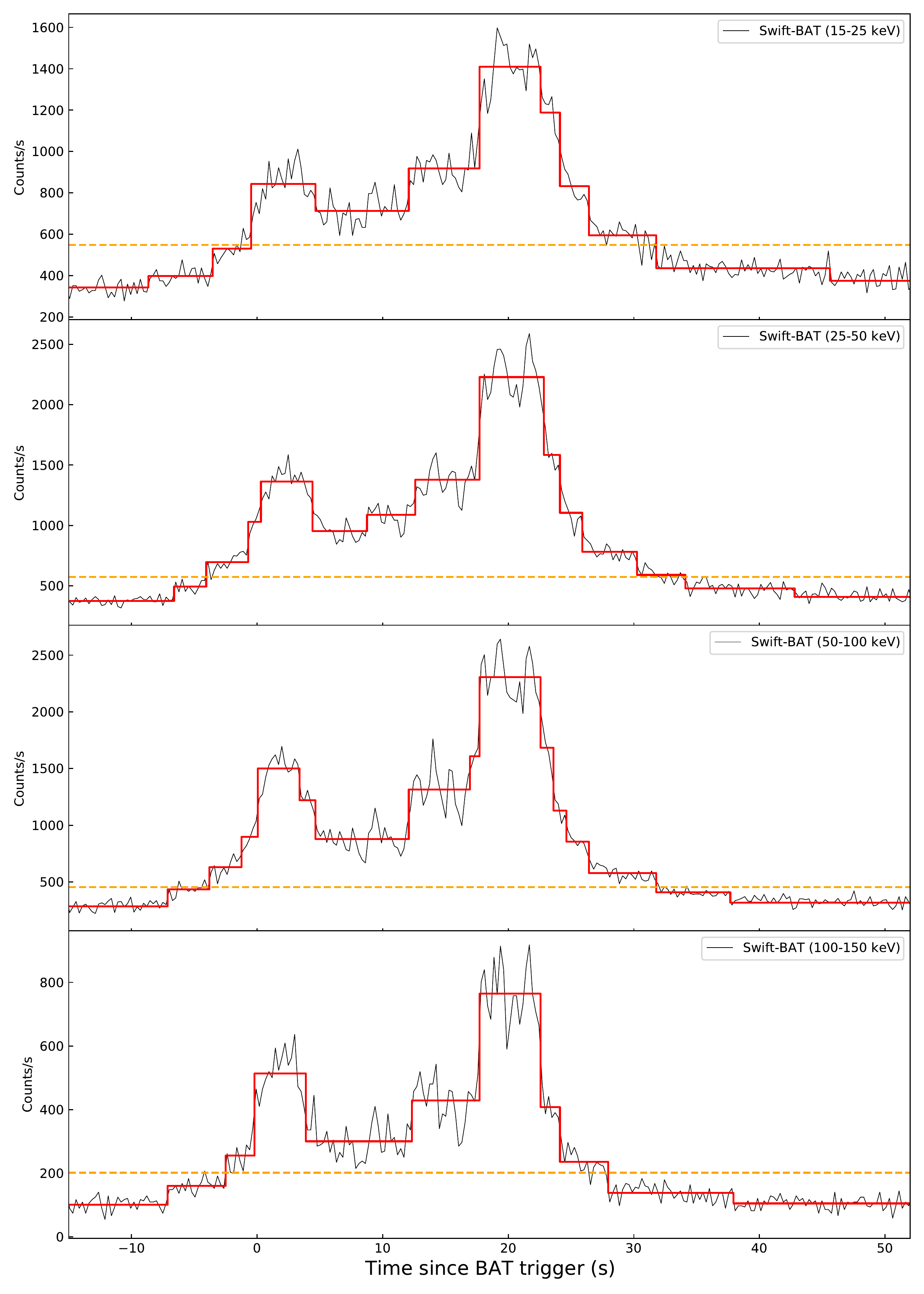}
\includegraphics[angle=0,width=0.45\textwidth]{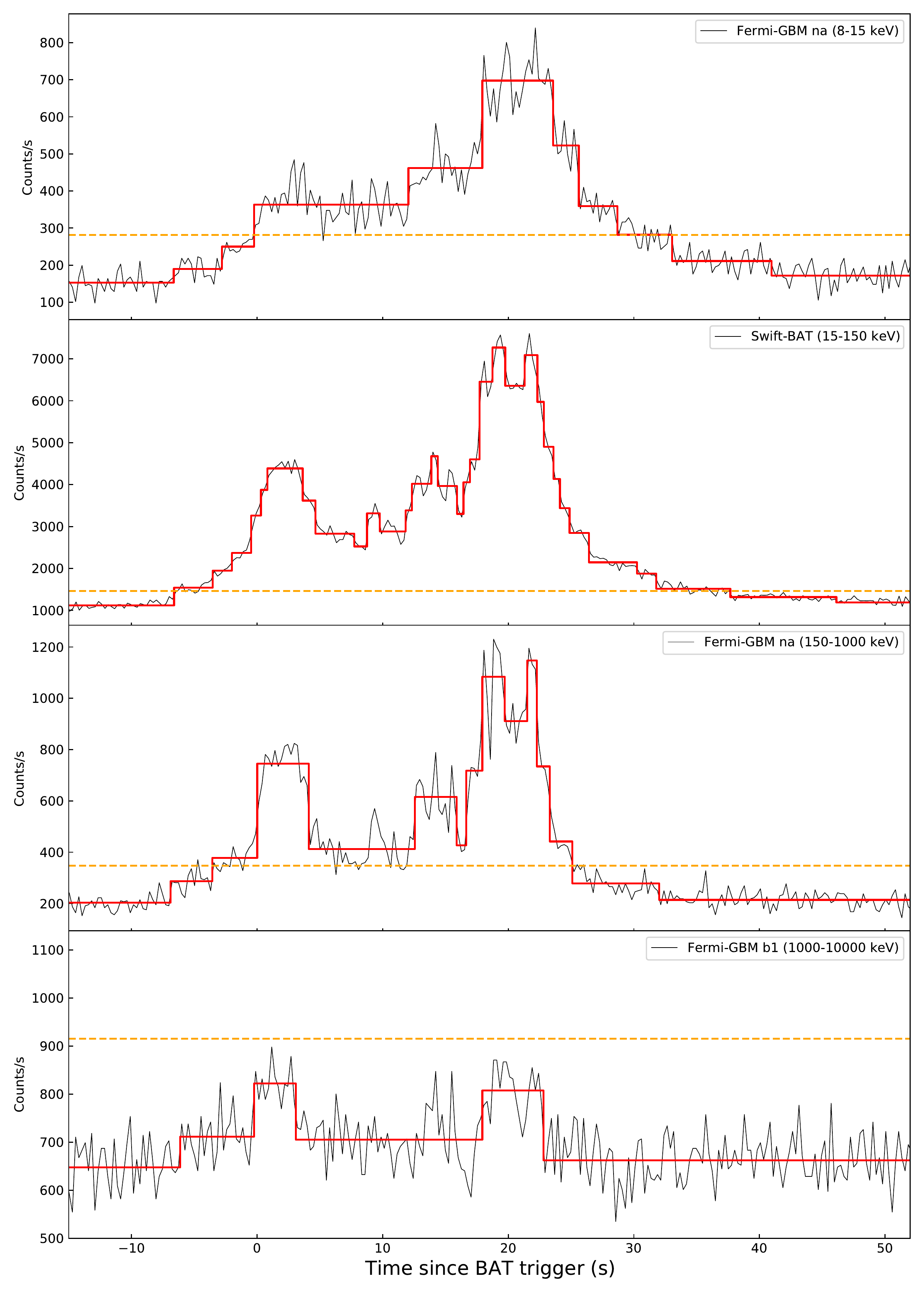}
\caption{Prompt gamma-ray lightcurves of GRB~201216C. The solid red step lines mark the analysis results with the Bayesian block algorithm. The dash orange horizontal lines represents $\rm SNR=10$.}
\label{prompt2}
\end{figure}

GRB~201216C was detected with Swift/BAT \citep{Beardmore-2020-Gropp-GCN.29061....1B}.
Its redshift is 1.1 \citep{Vielfaure-2020-Izzo-GCN.29077....1V}.
Similar to our analysis to GRB~201015A, we extract its BAT lightcurves in different energy bands,
as shown in Figure~\ref{prompt2}.
One can observe that its lightcurves are composed of overlapping pulses,
with a duration of $29.82\pm 1.39$~s in the $50-300$~keV band measured with the Bayesian block algorithm \citep{Scargle-2013-Norris-ApJ...764..167S}.

\begin{figure}[htbp]
\centering
\includegraphics[angle=0,width=0.45\textwidth]{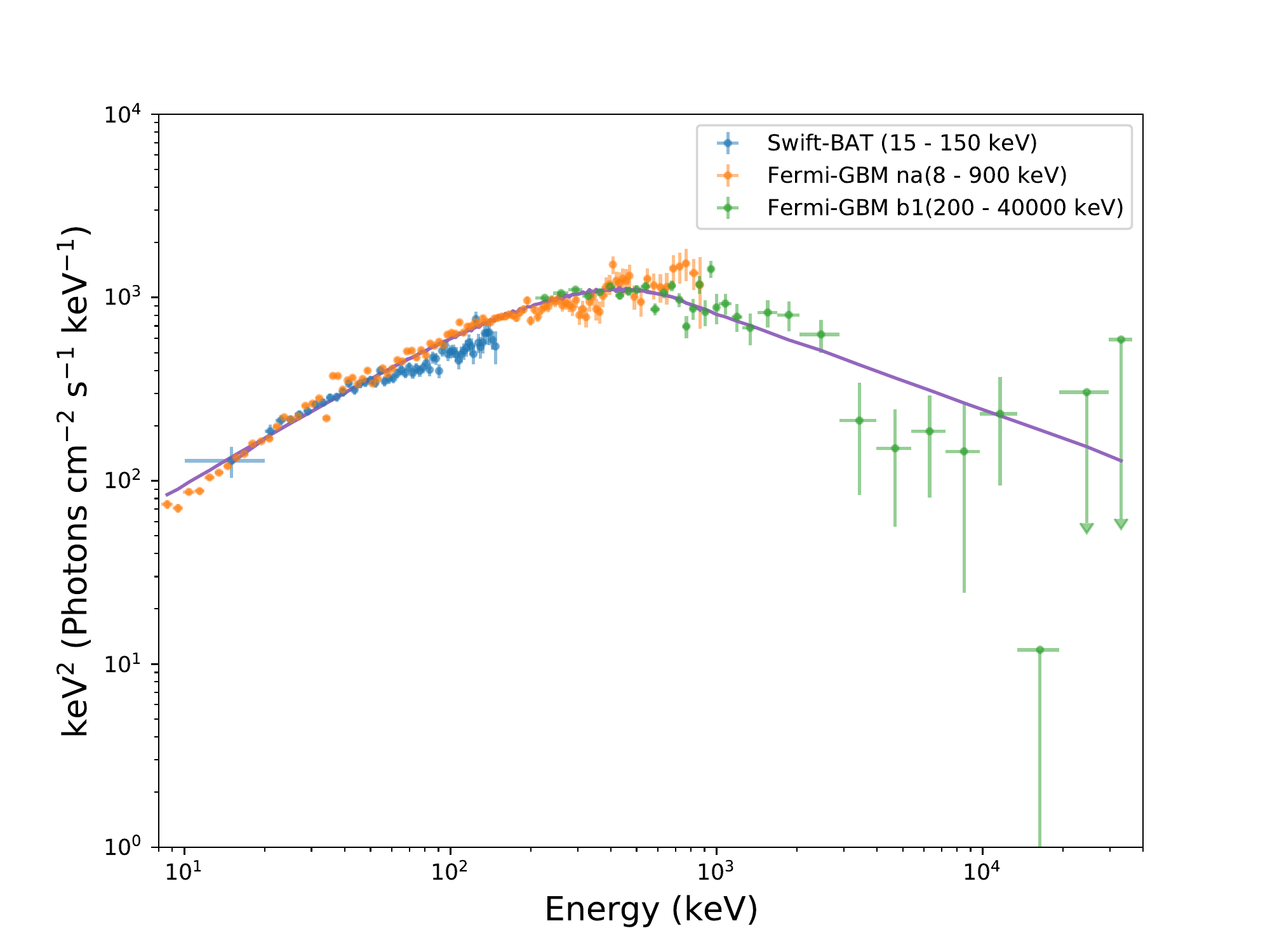}
\caption{Prompt gamma-ray spectra of GRB~201216C observed with Swift/BAT ($15-150$~keV), Fermi/GBM-NaI ($8-900$~keV), and Fermi/GBM-BGO ($200-40000$~keV). The purple solid line is our joint fit result with the Band function following the procedure described in \citet{Wang_2022_Jiang_arXiveprints_v.p2205..2982arXiv}. }
\label{grbm}
\end{figure}

GRB~201216C was also detected with {\em Fermi}/GBM \citep{Fermi_GBM_Team_2020GCN.29063....1F}.
It is particularly bright in the GRB band ($8-1000$~keV).
Its spectrum observed from $T_0+5.38$~s to $T_0+35.33$~s is well fitted with the Band function \citep{Band-1993-Matteson-ApJ...413..281B}, as shown in Figure~\ref{grbm}.
The derived peak energy of its $\nu f_\nu$ spectrum is $E_{p}=438_{-11}^{+12}~\rm$~keV and the photon indices are $\alpha=-1.12\pm 0.01$ and $\beta=-2.53_{-0.05}^{+0.04}$.
Its isotropic gamma-ray energy in the $1-10^4$~keV band is $E_{\rm \gamma,iso}= (5.76\pm 0.21) \times 10^{53}$~erg.

The X-ray afterglow of GRB~201216C was observed since $2966.8$~s after the BAT trigger \citep{Campana-2020-Cusumano-GCN.29064....1C}. We also download X-ray data from the web site of Swift Burst Analyzer\footnote{\url{https://www.swift.ac.uk/burst_analyser/01013243/}}.
The power-law decay slope is $\alpha_{\rm X}=1.74\pm 0.06$,
and the power-law photon index is $\Gamma_X=2.0\pm 0.14$ of the X-ray afterglows.
Its optical afterglow data are taken from  \cite{Shrestha-2020-Melandri-GCN.29085....1S},
\cite{Izzo-2020-Malesani-GCN.29066....1I},
and \cite{Ror_2022_Gupta_arXiv221110036R}. We convert the $r^{'}$ and C band data to the R band.
The Galactic extinction correction to the burst direction is $r^{\prime}_{\lambda}=0.126$ and $R=0.119$ \citep{Schlafly-2011-Finkbeiner-ApJ...737..103S}.
Its radio afterglows were detected by the {\em e}-MERLIN, VLA, and MeerKAT telescopes
in the frequency of 5~GHz, 10~GHz, and 1.3~GHz, respectively \citep{Rhodes-2022-van_der_Horst-MNRAS.513.1895R}.
We adopt the 10~GHz radio data for our analysis.
The multi-wavelength afterglow lightcurves of GRB~201216C are shown in the right panel of Figure~\ref{201015A}.

\section{Modeling the Radio-Optical-X-ray Afterglows}
\label{modeling}
\subsection{Model}
We attribute the afterglows to the emission from synchrotron plus SSC process of the electrons accelerated by the external shocks. The radiation physics and dynamics of the external shock fireball models please refer to \cite{Meszaros_1997_Rees_apj_v476.p232..237},
\cite{Sari-1998-Piran-ApJ...497L..17S},
\cite{Dermer_2000_Chiang_ApJ...537..785D},
\cite{Huang-2000-Gou-ApJ...543...90H},
\cite{Sari_2001_Esin_ApJ...548..787S},
\cite{Zhang_2001_Meszaros_ApJ...559..110Z},
\cite{Wang_2001_Lu_ApJ...556.1010W},
\cite{Fan-2006-Piran--MNRAS.369..197F},
and \cite{Ren-2020-Lin-ApJ...901L..26R,Ren_2022_Wang_arXiv221010673R}.
We outline our calculations as following.

We set the GRB jet as a uniform top-hat jet without considering the lateral expansion.
The surface of radiation is taken as a spherical shell with a radius $R(t,\theta)$ at a given time $t$, where $R$ is the radius of jet shell to the central engine and $\theta$ is the latitude angle respect to the jet axis.
Assuming the jet is viewed on axis,
the radiating surface is uniformly divided into 500 rings centering the jet axis in the logarithmic scale of the jet opening angle $\theta_j$.
The effect of the equal-arrival-time surface (EATS; e.g.,\citealp{Waxman_1997__apj_v491.p19..22L}) is considered by numerically calculating the arrival-time of photons from each ring.
The delay time of photons is $\Delta t(\theta)=R(\theta)(1-\cos \theta)/c$ for a ring at $\theta$ with respect to the axis.
Additionally, the Doppler factor of an on-axis viewed observer reads $D(\theta)=[\Gamma(1-\beta\cos{\theta})]^{-1}$,
where $\Gamma$ is the bulk Lorentz factor of jet and $\beta=\sqrt{1-\Gamma^{-2}}$.
The observed flux as a function of $\theta$ is $F(\theta)=D^3(\theta) F^{\prime}(\theta)$,
and the observed frequency is $\nu(\theta)=D(\theta)\nu^{\prime}(\theta)$,
where `$\prime$' represents the co-moving frame of the shocks.
The observed flux $F(t,\theta,\nu)$ is calculated by considering the Doppler boosting effect
and integrating over the EATS to get the observed SEDs.

It is assumed that the energy distribution of electron injection after
shocked blast wave is modeled as a single power-law function,
i.e., $ Q(\gamma^{\prime}_e)\propto {\gamma^{\prime}_e}^{-p} $ for
$ \gamma^{\prime}_{e,\min} \leqslant \gamma^{\prime}_{e} \leqslant \gamma^{\prime}_{e,\max}$.
We have $ \gamma^{\prime}_{e,\min}=\epsilon_e(p-2)m_{\rm p}\Gamma/[(p-1) m_{e}]$,
and $ \gamma^{\prime}_{e,\max}=\sqrt{9m_e^2 c^4/(8B^{\prime}q_{e}^3)}$,
among $\Gamma$, $c$ and $m_p$ is the Lorentz factor of the fireball,
the speed of light and the proton mass, $m_e$ and $q_e$ are the mass and charge of the electron, respectively.
It should be noted that $B^{\prime}=\sqrt{32\pi\epsilon_{B} n}\Gamma c$ represents the magnetic field,
$n$ is the number density of the medium.
Morever, $\epsilon_e$ and $\epsilon_B$ indicate the electron and magnetic field energy partition fractions of the fireball internal energy, respectively
(e.g., \citealp{Sari-1998-Piran-ApJ...497L..17S,Kumar_2012_Hernandez_mnras_v427.p40..44L}).
The radiation efficiency can be given by the equation $\epsilon = \epsilon_{\rm rad} \epsilon_e$
with $\epsilon_{\rm rad}=\min\{1,(\gamma_{e,\min}/\gamma_{e,c})^{(p-2)}\}$
\citep{Sari_2001_Esin_apj_v548.p787..799, Fan_2008_Piran_mnras_v384.p1483..1501}, where
$ \gamma_{e,c}=6 \pi m_{e} c /(\sigma_{\mathrm{T}} \Gamma {B^{\prime}}^2 t^{\prime}) $
is the efficient cooling Lorentz factor of electrons.
In the fast-cooling regime (${\gamma^{\prime}_{e,\min}}>{\gamma^{\prime}_{e,c}}$),
the instantaneous electron spectrum
$n^{\prime}_{e}(\gamma^{\prime}_{e}) $
can be derived as follows:
\begin{equation}
n^{\prime}_{e}(\gamma^{\prime}_{e}) = A_1
\left\{
\begin{array}{ll}
{\gamma^{\prime}_{e}}^{-p}, &{\gamma^{\prime}_{e,\min}} \leqslant \gamma^{\prime}_e<{\gamma^{\prime}_{e,c}},\\
{\gamma^{\prime}_{e,c}}{\gamma^{\prime}_{e}}^{-(p+1)},& {\gamma^{\prime}_{e,c}}\leqslant{\gamma^{\prime}_e}\leqslant {\gamma^{\prime}_{e,\max}},
\end{array}
\right.
\end{equation}
and in the slow-cooling regime (${\gamma^{\prime}_{e,\min}}<{\gamma^{\prime}_{e,c}}$),
\begin{equation}
n^{\prime}_e(\gamma^{\prime}_e)=A_2
\left\{
\begin{array}{ll}
{\gamma^{\prime}_e}^{-2}, & {\gamma^{\prime}_{e,c}} \leqslant \gamma^{\prime}_e < {\gamma^{\prime}_{e,\min}},\\
{\gamma^{\prime}_{e,\min}}^{p-1}{\gamma^{\prime}_e}^{-(p+1)},& {\gamma^{\prime}_{e,\min}} \leqslant \gamma^{\prime}_e \leqslant {\gamma^{\prime}_{e,\max}}.
\end{array}\right.
\end{equation}
Where $A_1=N_{e,\rm tot}\times p(1-p){({\gamma_c}^{1-p} - p{\gamma_{e,\min}}^{1-p})^{-1}}$, $A_2=N_{e,\rm tot}\times p[{(1-p)/{\gamma_{e,\min}}-p/{\gamma _{e,c}}}]^{-1}$,
$N_{e,\rm tot}=\int_{0}^{r}n_e {r^{\prime}}^2 dr^{\prime}$,
here $n_e$ is the number density of leptons in the burst frame.
The power of synchrotron radiation at a given frequency $\nu^{\prime}$ is
\begin{equation}
P^{\prime}\left(\nu^{\prime}\right)
=\frac{\sqrt{3} q_{e}^{3} B^{\prime}}{m_{e} c^{2}} \int_{0}^{\gamma_{\max }^{\prime}} F\left(\frac{\nu^{\prime}}{\nu_{c}^{\prime}}\right) n_{e}^{\prime}(\gamma_{e}^{\prime}) d\gamma_{e}^{\prime},
\end{equation}
where ${F(x)= x \int_{x}^{+\infty} K_{5/3}(k)dk}$ with $K_{5/3}(k)$ being the modified Bessel function of $5/3$ order and
${\nu_{c}^{\prime}=3 q_{e} B^{\prime} {\gamma_{e}^{\prime}}^ 2 /(4 \pi m_{e} c)}$.
The effects of synchrotron self-absorption
\citep{Granot_2002_Sari_apj_v568.p820..829,Warren_2022_Dainotti_apj_v924.p40..40}
is also considered in the calculations.

The typical synchrotron emission
frequency, the cooling frequency and the peak spectral flux are provided as follows (\citealp{Sari-1998-Piran-ApJ...497L..17S},  coefficients taken from \citealp{Yost_2003_Harrison_ApJ...597..459Y}):
\begin{eqnarray}
&\nu_m&  =  3.3 \times 10^{12}~{\rm Hz}~
\left(\frac{p-2}{p-1}\right)^2(1+z)^{1/2}\epsilon_{B,-2}^{1/2}
\epsilon_{e,-1}^{2}E_{k,52}^{1/2} t_d^{-3/2} \label{num},\\
&\nu_c&  =  6.3 \times 10^{15}~{\rm Hz}~
(1+z)^{-1/2} (1+Y)^{-2} \epsilon_{B,-2}^{-3/2}E_{k,52}^{-1/2} n^{-1}
t_d^{-1/2} \label{nuc},\\
&F_{\nu,\max}&  =  1.6~{\rm mJy}~
(1+z)D^{-2}_{28}\epsilon_{B,-2}^{1/2}E_{k,52}n^{1/2}~
\label{Fnumax},
\end{eqnarray}
where $t_d$ is the observer's time in unit of days,
\begin{equation}
Y=[-1+(1+4 \epsilon_{\rm rad}\eta \epsilon_e / \epsilon_B)^{1/2}]/2,
\label{Y}
\end{equation}
is the inverse-Compton (IC) parameter,
and $\eta \leq 1$ is a correction factor introduced by the
Klein-Nishina correction. Note that the SSC effect is significantly suppressed in the Klein-Nishina regime for the photons with energy $\nu > \nu_{\rm KN}$, where
\begin{eqnarray}
\nu_{\rm KN} & = & h^{-1} \Gamma m_e c^2 \gamma_{e,X}^{-1} (1+z)^{-1}
\nonumber \\
& \simeq & 2.4 \times 10^{15} ~{\rm Hz}~ (1+z)^{-3/4} E_{52}^{1/4}
\epsilon_{B,-2}^{1/4} t_d^{-3/4} \nu_{18}^{-1/2} ~,
\end{eqnarray}
where $h$ is the Planck's constant.
Numerically,
we have $\eta =\min\{1, (\nu_{\rm KN} / \nu_c)^{(3-p)/2}\}$ for slow cooling and $\eta = \min\{1, (\nu_{\rm KN} / \nu_m)^{1/2}\}$ for fast cooling,
where the factors $(\nu_{\rm KN} /\nu_c)^{(3-p)/2}$ and $(\nu_{\rm KN} / \nu_m)^{1/2}$ denote the fractions of the photon energy density that contributes to self-IC in the X-ray band in the slow and fast cooling regimes, respectively.
A calculation of the emission of the SSC process is based on the electron spectrum and seed photons from the synchrotron radiation.
In the numerical solution, the Klein-Nishina effect and the $\gamma\gamma$ annihilation effect have both been addressed
(e.g., \citealp{Gould_1967_Schreder_PhRv..155.1404G,Fan_2008_Piran_mnras_v384.p1483..1501, Nakar_2009_Ando_ApJ...703..675N,Murase_2011_Toma_apj_v732.p77..77,Geng_2018_Huang_apjs_v234.p3..3}).
The EBL absorption effect is taken into account for calculating the observed high energy photons at the Earth \citep{Dominguez_2011_Primack_mnras_v410.p2556..2578}.
In addition, we also include the extinction effect of the GRB host galaxy in our calculation by adopting the Mike Way extinction-law with $R_V=3.1$ \citep{Pei-1992-ApJ...395..130P,Fitzpatrick-1999-PASP..111...63F,Barbary_2016___v.p..}.

\subsection{Numerical Results}
We develop a Python-Fortran hybrid code for our calculations.
We fit the radio, optical, and X-ray band lightcurves of GRB~201015A and GRB~201216C with our models by utilizing the {\tt ultranest} and {\tt emcee} Python packages, respectively\footnote{The package {\tt emcee} is a Markov Chain Monte Carlo (MCMC) algorithm \citep{Foreman_2013_Hogg-PASP..125..306F},
and the {\tt ultranest} is implemented by the nested sampling Monte Carlo algorithm MLFriends
\citep{Buchner_2016__StatisticsandComputing_v26.p383..392,Buchner_2019__pasp_v131.p108005..108005}
available at \url{https://johannesbuchner.github.io/UltraNest/} \citep{Buchner_2021__TheJournalofOpenSourceSoftware_v6.p3001..3001}}.
The model parameters are the isotropic kinetic energy ($E_{\rm k, iso}$),
the initial bulk Lorentz factor ($\Gamma_0$), the medium number density ($n_0$), the electron energy distribution power-law index $p$,
$\epsilon_{e}$, and $\epsilon_{B}$,
the GRB host galaxy extinction parameter $E_{B-V}$
and the jet opening angle $\theta_{j}$.
Our fitting results are shown in Figure~\ref{201015A}.
The derived model parameters are reported in Table \ref{tab1}.

\begin{figure}[htbp]
\centering
\includegraphics[angle=0,width=0.9\textwidth]{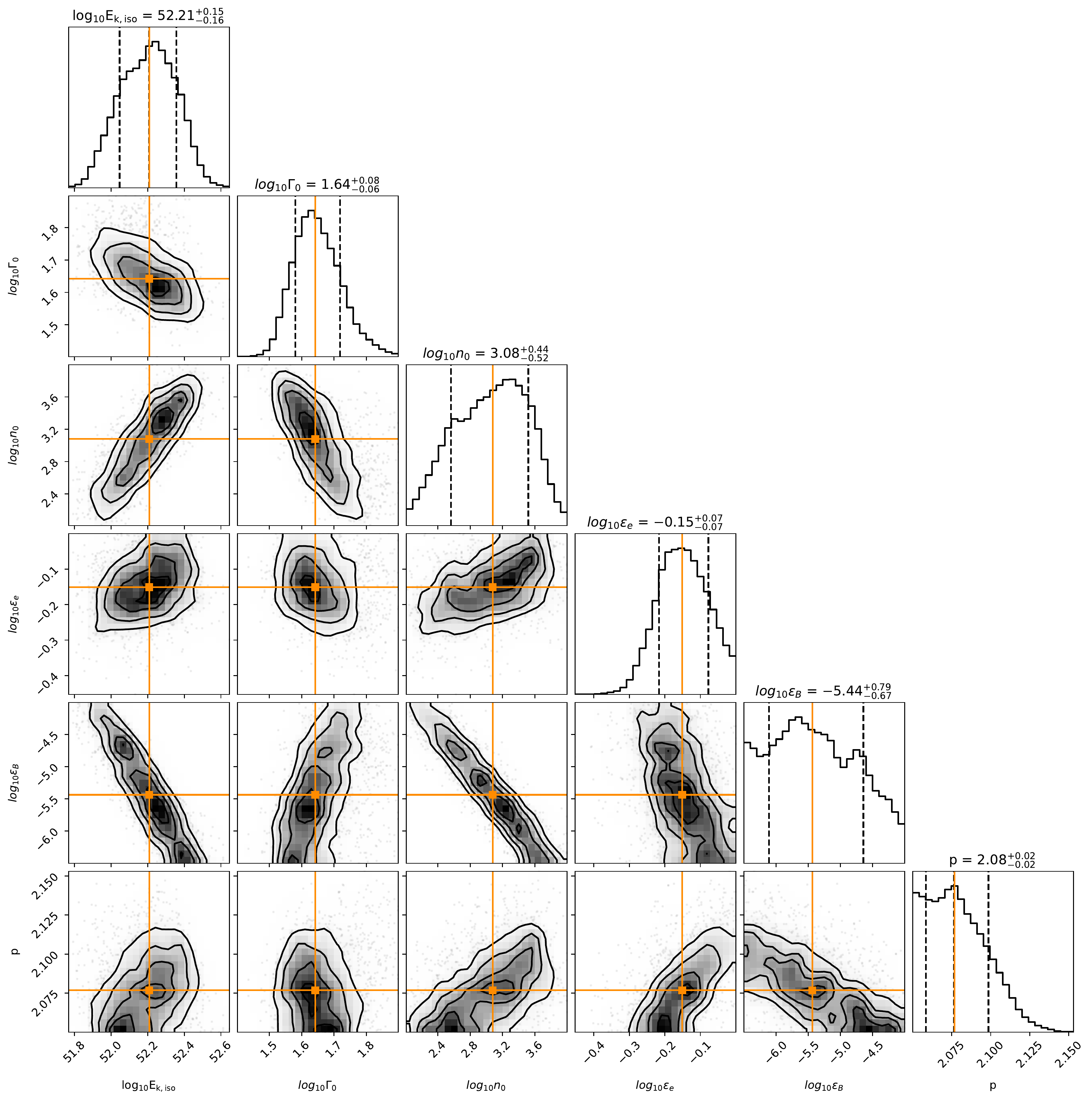}
\caption{Posterior distribution contours of the model parameters derived from the fit for GRB~201015A. The vertical dashed lines mark the $1\sigma$~confidence level regions centering at their median probabilities.}
\label{corner1}
\end{figure}

Figures~\ref{corner1} shows the probability distributions of the model parameters for GRB~201015A.
It is found that they are constrained well.
We have ${\rm log_{10}}\Gamma_{0}=1.64_{-0.06}^{+0.08}$,
indicating that its jet is moderate relativistic.
Its kinetic energy is also at the low end of typical GRBs,
i.e., ${\rm log_{10}}E_{\rm k, iso}({\rm erg}) =52.21_{-0.16}^{+0.15}$.
The electron distribution, the energy faction of magnetic field is consistent with typical GRBs, i.e., $p=2.08\pm 0.02$,
${\rm log_{10}}\epsilon_B=-5.44_{-0.67}^{+0.79}$.
However, the energy faction of electrons is larger more,
i.e., ${\rm log_{10}}\epsilon_e=-0.15\pm 0.07$,
The derived medium density is very large,
i.e., ${\rm log_{10}}n_0({\rm cm}^{-3})=3.08_{-0.52}^{+0.44}$,
but the host galaxy extinction for GRB~201015A is negligible.

\begin{figure}[htbp]
\centering
\includegraphics[angle=0,width=0.9\textwidth]{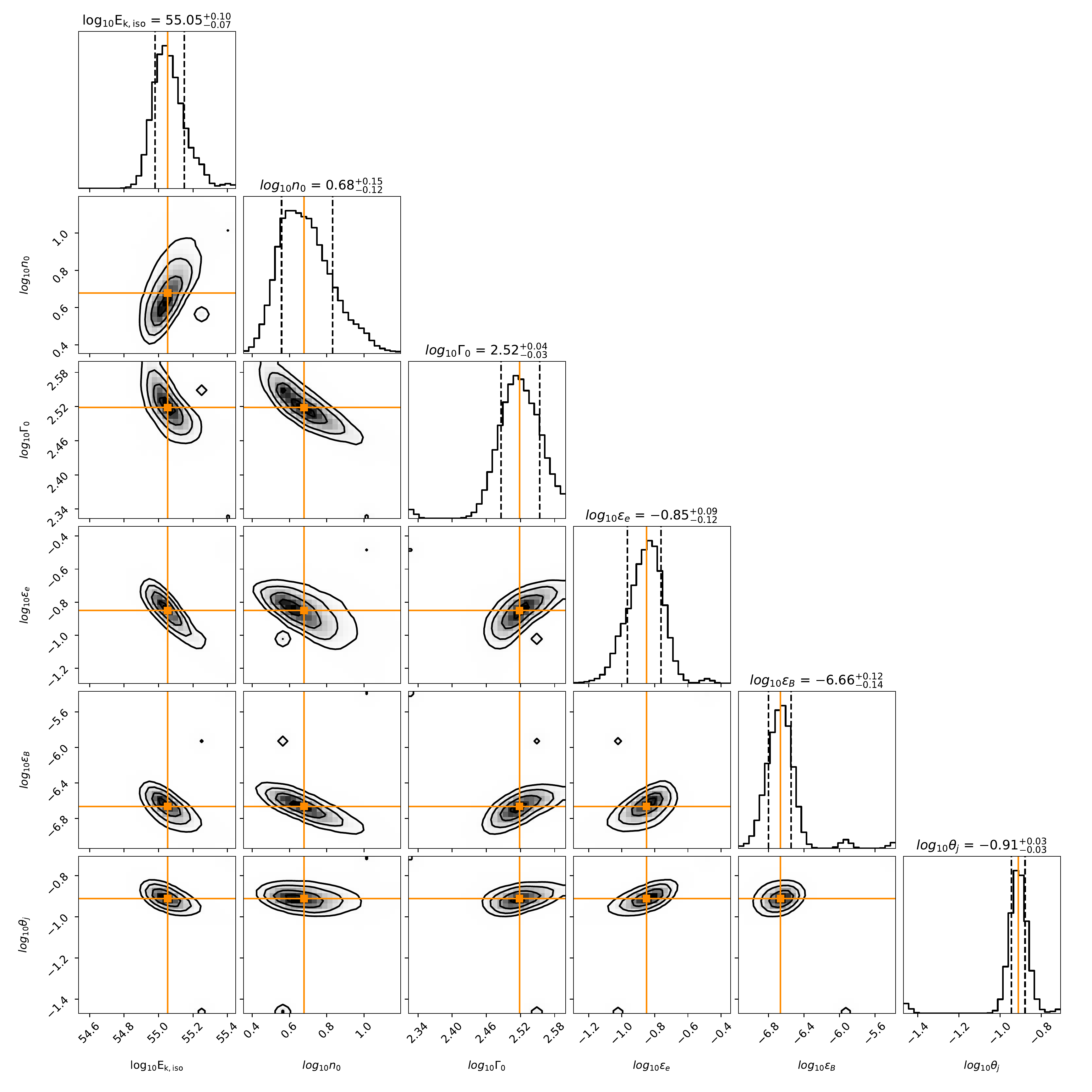}
\caption{Posterior distribution contours of the model parameters derived from our fits for GRB~201216C. The p=2.05 is fixed.}
\label{corner2}
\end{figure}

For the purpose of better limit the multi-band fitting of GRB 201216C afterglow lightcurves,
we assume that both the optical and X-ray afterglows in the same spectral regime of $\nu >\nu_c$,
hence $\rm p=2\beta_{X}$ (e.g., \citealp{Zhang_2006_Fan_ApJ...642..354Z}).
The photon index of the X-ray afterglows is $\Gamma_{\rm X}=2.0\pm 0.14$.
Therefore, we fix $p=2.05$. As presented in \cite{Rhodes-2022-van_der_Horst-MNRAS.513.1895R},
the environment of GRB 201216C is dusty.
Our joint spectral analysis of the X-ray and optical afterglows gives $E_{B-V}=1.26\pm 0.13$.
Fixing the values of $E_{B-V}$ and $p$,
we fit the mulitiband lightcurves of GRB~201216C using the {\tt emcee} Python package.
The upper limit of flux at $0.1-1$~GeV band observed by Fermi/LAT during $T_0+3500$~s to $T_0+5500$~s is also taken into account in our SED fit.
The derived probability distributions of the other model parameters are shown in the Figure \ref{corner2}.
We have ${\rm log_{10}}\Gamma_{0}=2.52^{+0.04}_{-0.03}$, indicating that its jet is ultra-relativistic.
Its kinetic energy is at the high end of typical GRBs, e.g., ${\log_{10}}E_{\rm k, iso}({\rm erg}) =55.05_{-0.07}^{+0.10}$.
The energy faction of electron is consistent with typical GRBs, e.g.,
${\rm log_{10}}\epsilon_e=-0.85^{+0.09}_{-0.12}$.
However, the portion of magnetic field is much smaller, e.g.,
$\rm log_{10}\epsilon_{B}=-6.66_{-0.14}^{+0.12}$.
The medium density of circumburst is moderately, e.g.,
$\rm log_{10} n_{0}({\rm cm}^{-3})=0.68_{-0.12}^{+0.15}$,
 and
$\rm log_{10}\theta_{j}(rad)=-0.91\pm 0.03$.

\section{Discussion}
\label{VHE ssc}
\subsection{VHE afterglows of GRBs 201015A and 201216C: model prediction {\em vs.} observations}
\begin{figure}[htbp]
 \centering
\includegraphics[angle=0,width=0.45\textwidth]{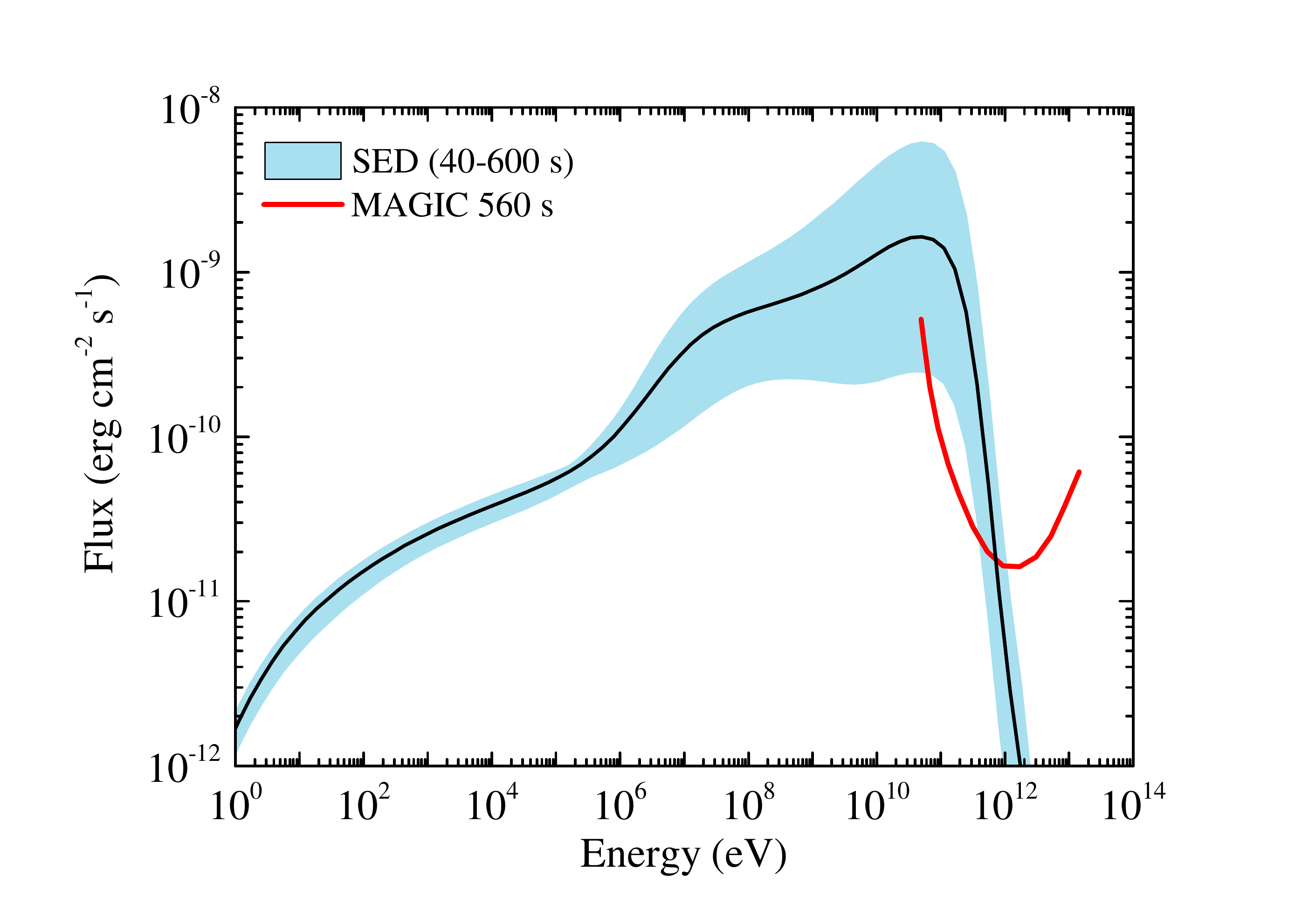}
\includegraphics[angle=0,width=0.45\textwidth]{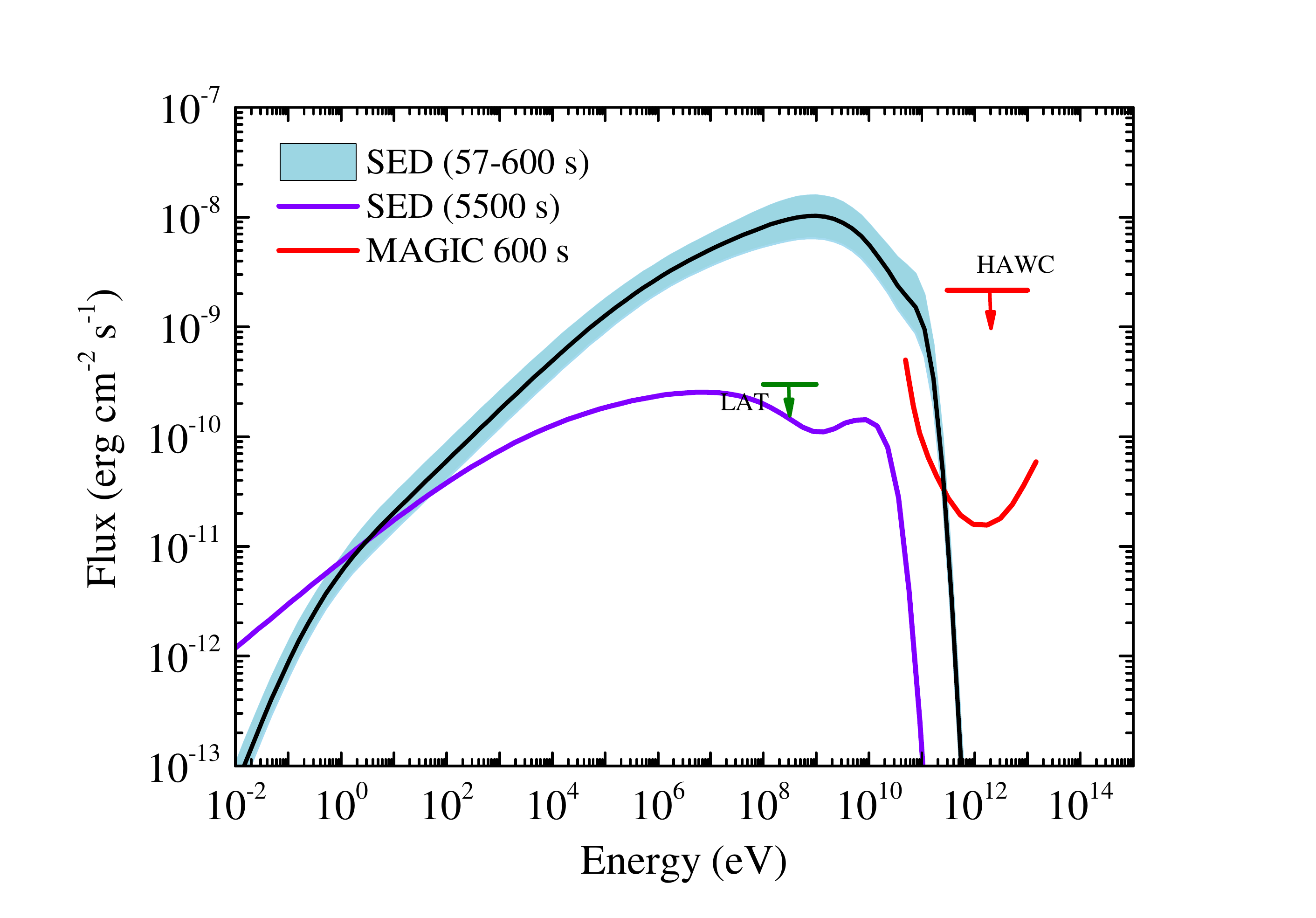}
\caption{Theoretically predicted afterglow SEDs of GRB~201015A at $40-600$~seconds (left panel) and GRB~201216C at $57-600$~seconds (right panel), in which emission is attributed to the synchrotron and SSC emissions of leptons. The shaded areas mark the uncertainties based on the errors of posterior distribution of parameters.
The red line represents the sensitivity of MAGIC for a given observation epoch, which is derived from the integral sensitivity for point-like sources with a Crab Nebula-like spectrum in 50 hours of observations \citep{Aleksic_2016_Ansoldi_APh....72...76A}.
The red and green down arrows represent the LAT and HAWC limits, respectively. The SED of GRB~201216C at 5500~seconds is also shown (purple line) to exam whether the model prediction violates the limit observed with Fermi/LAT.}
\label{SED}
\end{figure}

MAGIC collaboration announced the detection of VHE afterglows of GRBs~201015A and 201216C at $t>40$~s and $t>57$~s
with a confidence of $>3\sigma$ and $>5\sigma$, respectively \citep{Blanch-2020-Gaug-GCN.28659....1B, Blanch-2020-Longo-GCN.29075....1B}. The MAGIC team still does not release the VHE afterglow data so far. We examine whether the VHE afterglows inferred from above afterglow model coincide with the MAGIC observations in this section. We calculate their SEDs of the afterglows in the time slices $40-600$~s for GRB~201015A and $57-600$~s for GRB~201216C. The EBL absorption is taken into account in our calculations. The derived SEDs are shown in the Figure~\ref{SED}.

One can observe that the $>1$~MeV gamma-ray afterglows of GRB~201015A at the selected time slice are dominated by the SSC emission, which peaks at $\sim 50$~GeV. The gamma-ray afterglows are detected with MAGIC in the energy band from 60 to 600~GeV. The model-predicted flux in this range is $F_{\rm mo}=2.1\times10^{-9}~\rm erg~cm^{-2}~s^{-1}$. The threshold flux in this range is $F_{\rm th}=1.7\times 10^{-10}~\rm erg~cm^{-2}~s^{-1}$. We evaluate the detection significance of the predicted flux with a ratio of $R=(F_{\rm mo}-F_{\rm th})/{F_{\rm th}}$. It is found that the $R=11$. The team of the MAGIC telescopes reported detection of GRB~201015A at the energy band of $> 140~\rm GeV$ with a confidence level of $3.5\sigma$ \citep{Suda-2022-Artero-icrc.confE.797S}. Our model may over-estimate the VHE gamma-ray afterglow flux.

GRB 201216C is at the redshift of 1.1.
Although its VHE afterglows are suffered strong EBL absorption for its high redshift nature,
the MAGIC telescope team reported detection of GRB~201216C in the energy band around 100~GeV with a confidence level of $6\sigma$ \citep{Fukami-2022-Berti-icrc.confE.788F}.
The inferred SED of GRB 201216C from our model at the MAGIC observation epoch shows a very bright peak at $\sim 1$ GeV ($F_{\rm peak}\sim 1.0\times10^{-8}~\rm erg~cm^{-2}~s^{-1}$), which is dominated by the synchrotron radiations of the electrons. The gamma-ray flux in the MAGIC band (50-200 GeV) is dominated by the SSC component. The inferred detection confidence level with the MAGIC telescopes is $R=6.9$, which is comparable to that reported by \cite{Fukami-2022-Berti-icrc.confE.788F}.
The VHE afterglows of GRB~201216C were also observed by the HAWC telescope
in the $0.3-100$~TeV band since the GRB trigger to about one hour.
An upper limit flux at $95\%$ confidence level is obtained as
$dN/dE=4.05\times10^{-10}~E^{-2}~{\rm TeV}~{\rm cm}^{-2}~{\rm s}^{-1}$ \citep{Ayala-2020-GCN.29086....1A}. We calculated the upper limit by integrating the flux and obtained a limit of $2.16\times 10^{-9}~\rm erg~cm^{-2}~s^{-1}$ in the $0.3-100$~TeV band.
Our model prediction does not violate the limit.
In addition, the Fermi/LAT observed GRB~201216C from 3500 to 5500~s post the GBM trigger,
the upper limit of the energy flux in the energy band from 0.1 to 1~GeV is
$3\times10^{-10}~\rm erg~cm^{-2}~s^{-1}$ \citep{Bissaldi-2020-Omodei-GCN.29076....1B}.
We also plot the SED at 5500~s and mark the upper limit in the right panel of Figure~\ref{SED}.
It is found that our result also does not violate the limit.

\subsection{Diversity of GRBs with detection of VHE gamma-ray afterglows}
\label{discuss}
GRBs~201015A and 201216C, together with GRBs~180720B, 190829A and 190114C, would be representatives of the long GRBs with detection of VHE afterglows. The temporal and spectral properties of the prompt gamma-rays of the five GRBs are diverse.
They make a VHE-afterglow-selected GRB samples in broad ranges of burst energetic ($E_{\rm \gamma, iso}=10^{50}\sim 6\times10^{53}$~erg), peak energy of the $\nu f_\nu $ spectrum ($E_{\rm p}=14\sim 620$~keV), and redshift ($z=0.0785\sim 1.1$).
Except for GRB~201216C, the other three GRBs are associated with a supernova (SNe), i.e., GRB~201015A/AT 2020wyy \citep{Lipunov_2020_Gorbovskoy_TNSTR3144....1L}, GRB~190829A/SN~2019oyw \citep{Hu-2021-Castro-Tirado-A&A...646A..50H}, GRB~190114C/SN~2019jrj \citep{Melandri-2022-Izzo-A&A...659A..39M}.
GRB~201216C is a typical long GRB and the non-detection of its associated SN would be due to its high redshift nature since the current samples of SNe associated with GRBs are at $z=0.0085\sim 0.677$. The current highest-$z$ GRB/SN holder is GRB 111209A/SN 2011kl at $z=0.677$ \citep{Greiner_2015_Mazzali_Natur.523..189G}. We scale the flux of the $r^{\prime}$-band SNe lightcurve of GRB 111209A/SN 2011kl observed with the GROND telescope to $z=1.1$, as shown in the right panel of Figure \ref{201015A}. It is found that the peak magnitude of the SN at such a distance is below 24.2 magnitude, the limit of GROND telescope in an exposure time of 8 minutes \citep{Greiner_2008_Bornemann_PASP..120..405G.
} Different from the other four GRBs, the prompt gamma-rays of GRB~201015A is very short and soft, i.e., $E_p\sim 14$~keV and $T_{\rm 90}\sim 2$~s with an extended emission lasting about 6~s in the $15-25~{\rm KeV}$ band. This suggests that GRBs from collapsars may also be short duration (e.g., \citealp{Zhang-2009-Zhang-ApJ...703.1696Z, Xin-2011-Liang-MNRAS.410...27X}).

\begin{figure}[htbp]
\centering
\includegraphics[angle=0,width=0.9\textwidth]{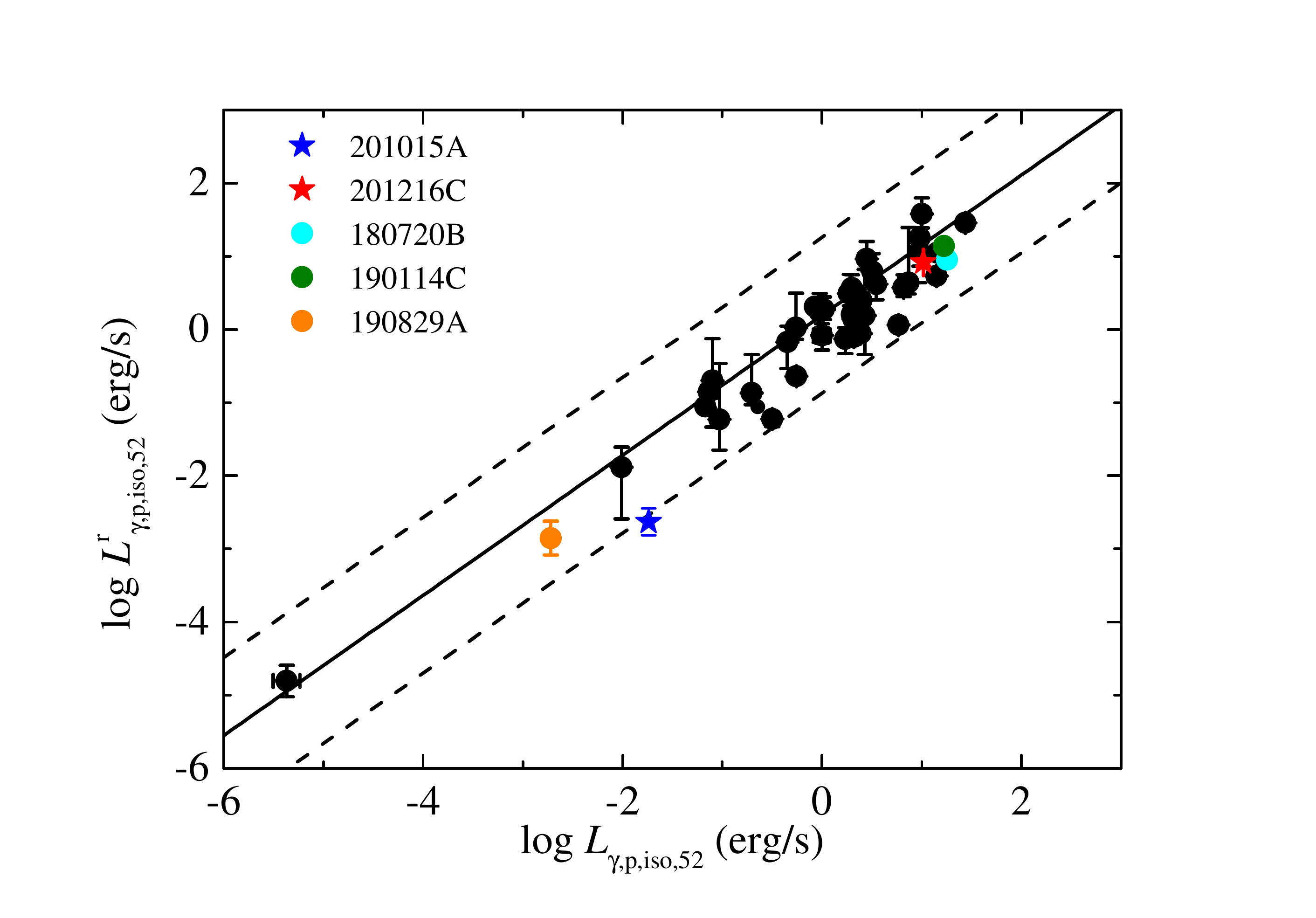}
\caption{GRBs with VHE gamma-ray afterglow detection in the $\log L^{r}_{\rm \gamma,p,iso}-\log L_{\rm \gamma,p,iso}$ plane in comparison with a sample of GRBs (black dots) including low-luminosity and high-luminosity GRBs from \citep{Liang_2015_Lin_ApJ...813..116L,Huang_2020_Liang_ApJ...903L..26H}, where the $\log L^{r}_{\rm \gamma,p,iso}$ is calculated with the relation $L_{\rm \gamma,p,iso}$--$\Gamma_{0}$--$E_{\rm p,z}$ \citep{Liang_2015_Lin_ApJ...813..116L}. The solid and dashed lines are the best fit and its dispersion in the $3\sigma$ confidence level for the relation between $\log L^{r}_{\rm \gamma,p,iso}$ and $\log L_{\rm \gamma,p,iso}$.}
\label{relation}
\end{figure}

GRBs~180720B, 190114C, and 201216C are very bright and have a very large Lorentz factor, i.e., $E_{\rm \gamma, iso}\sim 6\times 10^{53  }$~erg and $\Gamma_0=300-331$ \citep{Wang_2019_Liu_ApJ...884..117W}. GRBs~201015A and 190829A, which have $E_{\rm k,iso}\sim 10^{51}-10^{52}$~erg and $\Gamma_0\sim 40$, bridge the sub-energetic and energetic GRBs. Detection of VHE gamma-ray afterglows of GRBs~201216C and 201015A confirms that VHE gamma-ray afterglows can be detectable for both the ultra-relativistic, extremely-energetic GRBs and middle-relativistic, sub/moderate-energetic GRBs. It is found that they follow the $L_{\rm \gamma,p, iso}-E_{\rm p,z}-\Gamma_0$ relation that derived from typical GRBs, as shown in Figure~\ref{relation} \citep{Liang_2015_Lin_ApJ...813..116L}. GRB~201216C, together with GRBs 180720B and 190114C \citep{Huang_2020_Liang_ApJ...903L..26H}, is at the highest luminosity end, and GRBs~201015A and 190829A are in the lowest luminosity end of this relation for the current typical GRB samples.

The prompt gamma-ray radiation efficiency,
which is defined as $\eta_\gamma=E_{\rm \gamma, iso}/(E_{\rm \gamma, iso}+E_{\rm k, iso})$,
is a critical property of the radiating fireball. \cite{Beniamini-2015-Nava-MNRAS.454.1073B} calculated the GRB efficiency using the GeV afterglow data observed with the Fermi/LAT. They showed that $\eta_{\gamma}=1.5\%\sim 30\%$ for some typical GRBs. Based on our above analysis, we obtain  $\eta_\gamma\simeq 4.88\%$ for GRB 201216C and $\eta_\gamma\simeq 0.67\%$ for GRB~201015A. The $\eta_\gamma$ values of the two GRBs are much lower than that of the other GRBs with detection of sub-TeV/TeV gamma-ray afterglows, i.e.,
GRB 190829A ($16.4\%$; \citealp{Zhang-2021-Ren-ApJ...917...95Z}), GRB 180720B ($\eta_{\gamma}=37.5\%$), and GRB 190114C ($\eta_{\gamma}=33.3\%$) by adopting the $E_{\rm k}$ values from \cite{Wang_2019_Liu_ApJ...884..117W}.

\begin{figure}[htbp]
\centering
\includegraphics[angle=0,width=0.45\textwidth]{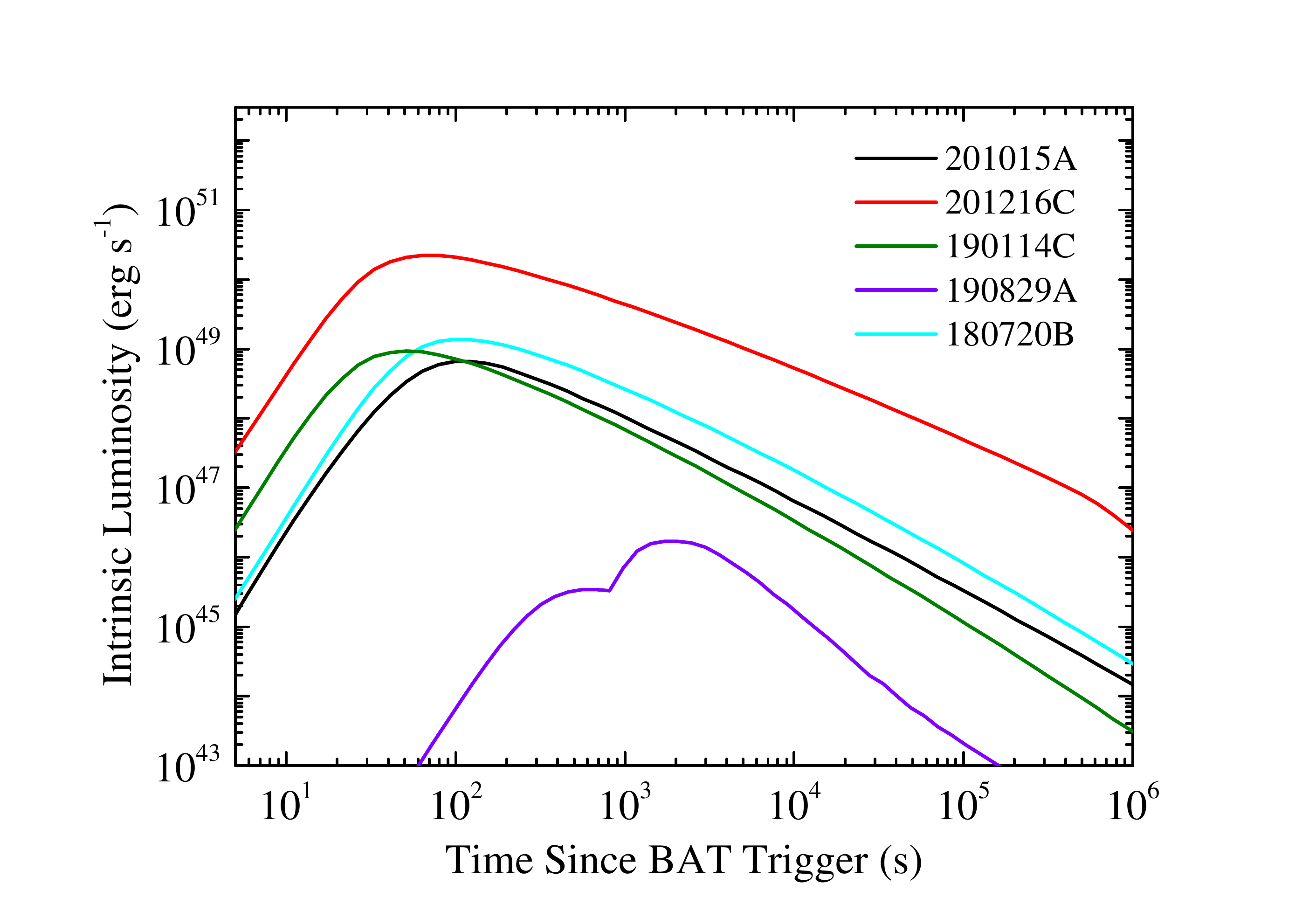}
\includegraphics[angle=0,width=0.45\textwidth]{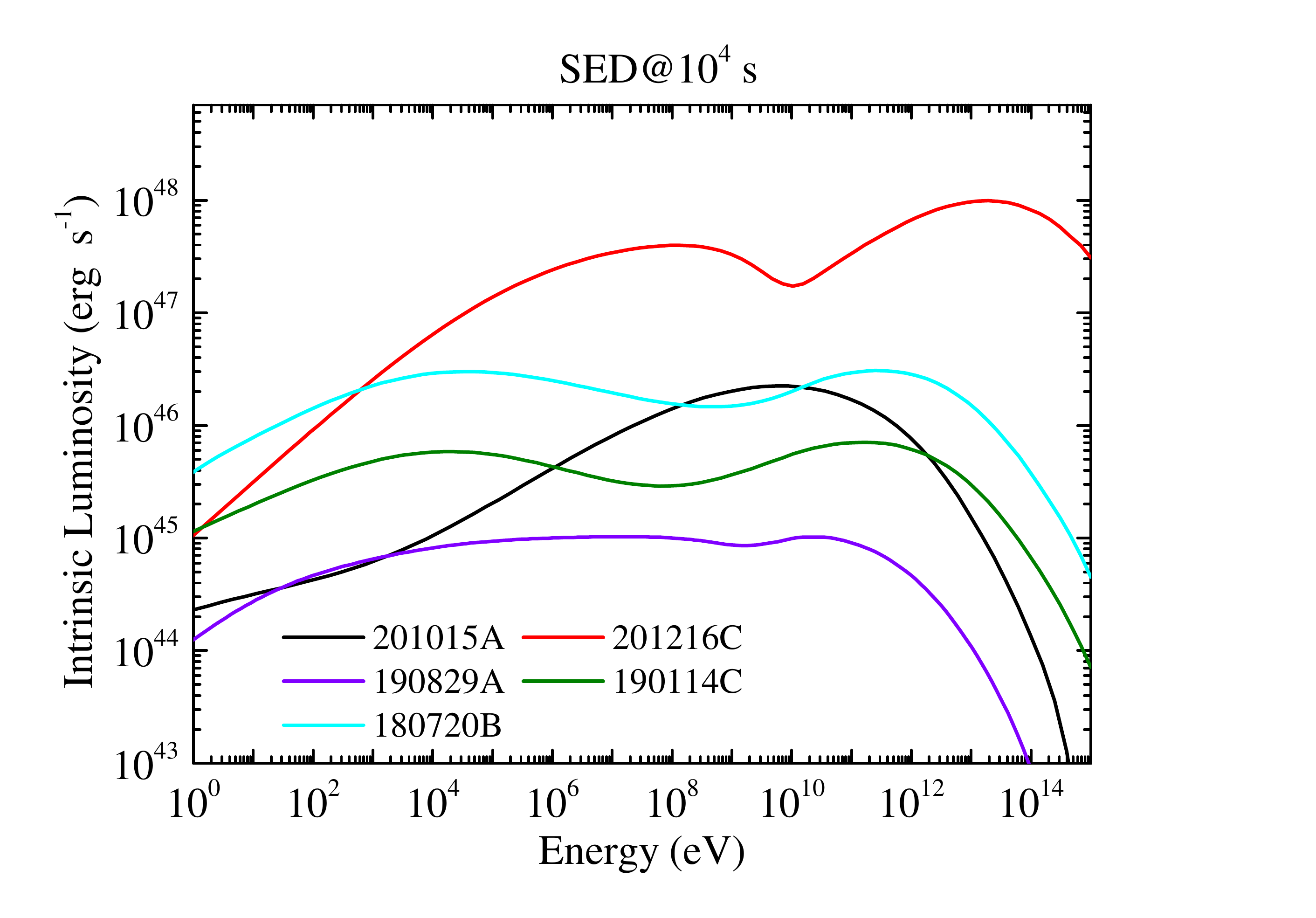}
\caption{The intrinsic luminosity lightcurves of GeV-TeV afterglows (left panel) and SEDs at $10^4$~seconds (right panel) of GRBs~201015A, 201216C, 190829A, 190114C and 180720B \citep{Wang_2019_Liu_ApJ...884..117W}.
In the left panel, we compare the intrinsic luminosity without EBL absorption effects. In the right panel, we compare five GRBs of VHE emissions at $10^4$~seconds without EBL absorption at same distance ($z=0.0785$).}
\label{LC}
\end{figure}

We compare the intrinsic VHE gamma-ray afterglow lightcurves and broadband SEDs at $10^{4}$~seconds of the five GRBs in Figure~\ref{LC}.
One can observe that GRB~201216C is the brightest one among them.
Its intrinsic SSC peaks at $E_{\rm p, SSC}\sim 20$ TeV with a specific luminosity of $L_{\rm p, SSC}\sim 5\times 10^{48}~\rm erg~s^{-1}$ at $t=10^{4}$~s.
Its $E_{\rm p, SSC}$ and $L_{\rm p, SSC}$ are higher than that of the other two energetic GRBs 180720B and 190114C by $1\sim 2$ orders of magnitude. Note that the $\Gamma_{0}$ values of the three GRBs are comparable, i.e., $\sim 330$, but the $E_{\rm k, iso}$ value of GRB 201216C is $1.12\times 10^{55}$ erg, being more than 1 orders of magnitude larger than the other two GRBs ($1\times 10^{54}$ erg for GRB 180720B and $6\times 10^{53}$ erg for GRB 190114C, \citealp{Wang_2019_Liu_ApJ...884..117W}).
Therefore, GRB 201216C should be intrinsically an ultra-bright event, being similar to GRB~221009A (e.g., \citealp{Ren_2022_Wang_arXiv221010673R,OConnor_2023_Troja_arXiv230207906O}). Its jet should be extremely narrow regarding the energetic crisis of the GRB phenomenon (e.g., \citealp{Racusin_2008_Karpov_Natur.455..183R}).
Based on photometry of the $g$, $r$, and $z$ bands,
\cite{Vielfaure-2020-Izzo-GCN.29077....1V} reported a spectral slope of  $\beta = 4.1\pm 0.2$ for GRB~201216C. This unusually red value suggests a significant extinction around the jet and its host galaxy, likely giving some insights to the nature of its progenitor. The intrinsic peak luminosity of $0.2-4$~TeV gamma-ray afterglows of GRB~190829A, a typical nearby low-luminosity GRB, is only $ 10^{45}$~erg/s at $t=10^4$ seconds, being lower than that of GRB~201216C by $\sim 4$ orders of magnitude.

The peak luminosity of $0.2-4$~TeV gamma-ray afterglow lightcurve of GRB~201015A is comparable to GRBs~180720B and 190114C, although the $E_{\rm k, iso}$ of GRB~201015A is lower than GRBs~180720B and 190114C by about 2 orders of magnitude. This is due to the SSC process dominates the cooling of the electrons in the jet of GRB~201015A. We obtain a Compton parameter $Y$ of 442 from Eq. \ref{Y}, adopting $\epsilon_{\rm rad}=1$ and $\eta_e=1$. The value of the $Y$ parameter is  larger than that of GRBs 190114C and 180720B by a factor of $10 \sim 14$.

The redshifts range of the five GRBs is from 0.0785 to 1.1. GRB~201216C is at the highest redshift ($z=1.1$) among the current GRBs with detection of sub-TeV/TeV gamma-ray afterglows, confirming that sub-TeV or TeV gamma-ray afterglows can be detectable with MAGIC beyond a redshift of 1.1, although they suffer strong EBL absorption. Middle-relativistic, sub/moderate-energetic GRBs would be preferred to be detected at low redshift. Marginal detection of VHE gamma-ray afterglows of GRB~201015A at a redshift comparable to GRB~190114C with the MAGIC telescope is due to occasionally catching its early at $\sim 60$~seconds. The solid detection of GRB~190829A up to 4~TeV at a very late epoch ($\sim 1$~day) is due to its nearby nature. TeV gamma-ray afterglows of GRBs would be promising targets for the VHE gamma-ray telescopes.

\section{Summary}
\label{summary}
We have analysed the prompt gamma-ray data of GRBs~201015A and 201216C and fit their radio-optical-X-ray afterglow lightcurves with the standard external-forward shock model by attributing the afterglows to the synchrotron emission from electrons accelerated by the shocks, and infer their VHE afterglows from the SSC scatterings. Our results are summarized  as following.

\begin{itemize}
    \item GRB~201015A is an extremely soft (with $E_p\sim 14$~keV) and short (with a sharp pulse lasting $\sim 2$~s followed by extended emission up to $\sim 6$~seconds in the $15-25~{\rm KeV}$ band) GRB from death of a massive star. It's jet is mid-relativistic (initial Lorentz factor $\Gamma_0=43.7$) surrounded by dense medium (with a density $n_0=1202.3~\rm cm^{-3}$). The inferred sub-TeV gamma-ray afterglows in the range from 60 to 600 GeV are detectable with the MAGIC telescopes, being roughly consistent with the observations.
    \item GRB~201216C is an extremely bright ($E_{\rm \gamma, iso}=5.8\times 10^{53}$~$\rm erg$) with a hard spectrum ($E_p=438~\rm keV$). It's jet is ultra-relativistic (initial Lorentz factor $\Gamma_0=331$) surrounded by dense and dusty medium ($n=4.8~\rm cm^{-3}$ and $E_{\rm B-V}=1.26\pm 0.13$). The inferred VHE gamma-ray afterglows is very bright at the early with a peak flux of $\sim 10^{-9}~\rm erg~cm^{-2}~s^{-1}$ in 60~GeV at $57-600$~seconds post the GRB trigger, making they can be detectable with the MAGIC telescopes in a high confidence level, although the GRB is at a redshift of 1.1.

    \item GRBs~201015A and 201216C, together with GRBs~180720B, 190114C and 190829A, are representatives of VHE gamma-ray afterglow selected long GRBs based on the diversity of their kinetic energy ($10^{51}\sim 10^{55}$~erg), medium density ($2\sim 1202~\rm cm^{-3}$), peak energy $E_p$ of the prompt gamma-ray spectrum $\nu f_\nu$, and the initial Lorentz factor ($30\sim 331$). The inferred peak energy of the intrinsic SSC component at $10^{4}$~seconds is around $0.03\sim 20$~TeV for both energetic (GRBs~180720B, 190114C, and 201216C) and sub-energetic (GRBs~190829A and 201015A) GRBs, but the flux level of the VHE gamma-ray afterglow is sensitive to the density of the ambient medium. VHE gamma-ray afterglows of GRBs would be promising targets for the TeV/sub-TeV telescopes.
\end{itemize}

\acknowledgments
We very appreciate helpful comments from the referee. We thank thoughtful discussion with Hou-Jun LV and Da-Bin Lin. We acknowledge the use of the Fermi archive’s public data. We acknowledge the use of the public data from the Swift data archive and the UK Swift Science Data Center.
This work is supported by the National Natural Science Foundation of China
(Grant No.12133003).

\software{\texttt{emcee}\citep{Foreman_2013_Hogg-PASP..125..306F},
\texttt{corner}\citep{ForemanMackey_2016__TheJournalofOpenSourceSoftware_v1.p24..24},
\texttt{extinction}\citep{Barbary_2016___v.p..},
\texttt{dynesty}\citep{Speagle_2020__mnras_v493.p3132..3158},
\texttt{ultranest}\citep{Buchner_2021__TheJournalofOpenSourceSoftware_v6.p3001..3001}
}

\clearpage


\begin{thebibliography}{}
\expandafter\ifx\csname natexlab\endcsname\relax\def\natexlab#1{#1}\fi
\providecommand{\url}[1]{\href{#1}{#1}}
\providecommand{\dodoi}[1]{doi:~\href{http://doi.org/#1}{\nolinkurl{#1}}}
\providecommand{\doeprint}[1]{\href{http://ascl.net/#1}{\nolinkurl{http://ascl.net/#1}}}
\providecommand{\doarXiv}[1]{\href{https://arxiv.org/abs/#1}{\nolinkurl{https://arxiv.org/abs/#1}}}

\bibitem[{{Abdalla} {et~al.}(2019){Abdalla}, {Adam}, {Aharonian}, {Ait
  Benkhali}, {Ang{\"u}ner}, {Arakawa}, {Arcaro}, {Armand}, {Ashkar}, {Backes},
  {Barbosa Martins}, {Barnard}, {Becherini}, {Berge}, {Bernl{\"o}hr},
  {Bissaldi}, {Blackwell}, {B{\"o}ttcher}, {Boisson}, {Bolmont}, {Bonnefoy},
  {Bregeon}, {Breuhaus}, {Brun}, {Brun}, {Bryan}, {B{\"u}chele}, {Bulik},
  {Bylund}, {Capasso}, {Caroff}, {Carosi}, {Casanova}, {Cerruti}, {Chand},
  {Chandra}, {Chen}, {Colafrancesco}, {Cury{\l}o}, {Davids}, {Deil}, {Devin},
  {deWilt}, {Dirson}, {Djannati-Ata{\"\i}}, {Dmytriiev}, {Donath},
  {Doroshenko}, {Dyks}, {Egberts}, {Emery}, {Ernenwein}, {Eschbach}, {Feijen},
  {Fegan}, {Fiasson}, {Fontaine}, {Funk}, {F{\"u}{\ss}ling}, {Gabici},
  {Gallant}, {Gat{\'e}}, {Giavitto}, {Giunti}, {Glawion}, {Glicenstein},
  {Gottschall}, {Grondin}, {Hahn}, {Haupt}, {Heinzelmann}, {Henri}, {Hermann},
  {Hinton}, {Hofmann}, {Hoischen}, {Holch}, {Holler}, {Horns}, {Huber},
  {Iwasaki}, {Jamrozy}, {Jankowsky}, {Jankowsky}, {Jardin-Blicq},
  {Jung-Richardt}, {Kastendieck}, {Katarzy{\'n}ski}, {Katsuragawa}, {Katz},
  {Khangulyan}, {Kh{\'e}lifi}, {King}, {Klepser}, {Klu{\'z}niak}, {Komin},
  {Kosack}, {Kostunin}, {Kreter}, {Lamanna}, {Lemi{\`e}re}, {Lemoine-Goumard},
  {Lenain}, {Leser}, {Levy}, {Lohse}, {Lypova}, {Mackey}, {Majumdar},
  {Malyshev}, {Marandon}, {Marcowith}, {Mares}, {Mariaud}, {Mart{\'\i}-Devesa},
  {Marx}, {Maurin}, {Meintjes}, {Mitchell}, {Moderski}, {Mohamed}, {Mohrmann},
  {Moore}, {Moulin}, {Muller}, {Murach}, {Nakashima}, {de Naurois},
  {Ndiyavala}, {Niederwanger}, {Niemiec}, {Oakes}, {O'Brien}, {Odaka}, {Ohm},
  {de Ona Wilhelmi}, {Ostrowski}, {Oya}, {Panter}, {Parsons}, {Perennes},
  {Petrucci}, {Peyaud}, {Piel}, {Pita}, {Poireau}, {Priyana Noel}, {Prokhorov},
  {Prokoph}, {P{\"u}hlhofer}, {Punch}, {Quirrenbach}, {Raab}, {Rauth},
  {Reimer}, {Reimer}, {Remy}, {Renaud}, {Rieger}, {Rinchiuso}, {Romoli},
  {Rowell}, {Rudak}, {Ruiz-Velasco}, {Sahakian}, {Sailer}, {Saito}, {Sanchez},
  {Santangelo}, {Sasaki}, {Schlickeiser}, {Sch{\"u}ssler}, {Schulz}, {Schutte},
  {Schwanke}, {Schwemmer}, {Seglar-Arroyo}, {Senniappan}, {Seyffert}, {Shafi},
  {Shiningayamwe}, {Simoni}, {Sinha}, {Sol}, {Specovius}, {Spir-Jacob},
  {Stawarz}, {Steenkamp}, {Stegmann}, {Steppa}, {Takahashi}, {Tavernier},
  {Taylor}, {Terrier}, {Tiziani}, {Tluczykont}, {Trichard}, {Tsirou}, {Tsuji},
  {Tuffs}, {Uchiyama}, {van der Walt}, {van Eldik}, {van Rensburg}, {van
  Soelen}, {Vasileiadis}, {Veh}, {Venter}, {Vincent}, {Vink}, {V{\"o}lk},
  {Vuillaume}, {Wadiasingh}, {Wagner}, {White}, {Wierzcholska}, {Yang},
  {Yoneda}, {Zacharias}, {Zanin}, {Zdziarski}, {Zech}, {Ziegler}, {Zorn},
  {{\.Z}ywucka}, {de Palma}, {Axelsson}, \&
  {Roberts}}]{Abdalla-2019-Adam-Natur.575..464A}
{Abdalla}, H., {Adam}, R., {Aharonian}, F., {et~al.} 2019, \nat, 575, 464,
  \dodoi{10.1038/s41586-019-1743-9}

\bibitem[{{Acciari} {et~al.}(2021){Acciari}, {Ansoldi}, {Antonelli}, {Arbet
  Engels}, {Asano}, {Baack}, {Babi{\'c}}, {Baquero}, {Barres de Almeida},
  {Barrio}, {Becerra Gonz{\'a}lez}, {Bednarek}, {Bellizzi}, {Bernardini},
  {Bernardos}, {Berti}, {Besenrieder}, {Bhattacharyya}, {Bigongiari}, {Biland},
  {Blanch}, {Bonnoli}, {Bo{\v{s}}njak}, {Busetto}, {Carosi}, {Ceribella},
  {Cerruti}, {Chai}, {Chilingarian}, {Cikota}, {Colak}, {Colombo}, {Contreras},
  {Cortina}, {Covino}, {D'Amico}, {D'Elia}, {da Vela}, {Dazzi}, {de Angelis},
  {de Lotto}, {Delfino}, {Delgado}, {Delgado Mendez}, {Depaoli}, {di Pierro},
  {di Venere}, {Souto Espi{\~n}eira}, {Dominis Prester}, {Donini}, {Dorner},
  {Doro}, {Elsaesser}, {Fallah Ramazani}, {Fattorini}, {Ferrara}, {Foffano},
  {Fonseca}, {Font}, {Fruck}, {Fukami}, {Garc{\'\i}a L{\'o}pez}, {Garczarczyk},
  {Gasparyan}, {Gaug}, {Giglietto}, {Giordano}, {Gliwny}, {Godinovi{\'c}},
  {Green}, {Green}, {Hadasch}, {Hahn}, {Heckmann}, {Herrera}, {Hoang},
  {Hrupec}, {H{\"u}tten}, {Inada}, {Inoue}, {Ishio}, {Iwamura}, {Jormanainen},
  {Jouvin}, {Kajiwara}, {Karjalainen}, {Kerszberg}, {Kobayashi}, {Kubo},
  {Kushida}, {Lamastra}, {Lelas}, {Leone}, {Lindfors}, {Lombardi}, {Longo},
  {L{\'o}pez-Coto}, {L{\'o}pez-Moya}, {L{\'o}pez-Oramas}, {Loporchio}, {Machado
  de Oliveira Fraga}, {Maggio}, {Majumdar}, {Makariev}, {Mallamaci}, {Maneva},
  {Manganaro}, {Mannheim}, {Maraschi}, {Mariotti}, {Mart{\'\i}nez}, {Mazin},
  {Mender}, {Mi{\'c}anovi{\'c}}, {Miceli}, {Miener}, {Minev}, {Miranda},
  {Mirzoyan}, {Molina}, {Moralejo}, {Morcuende}, {Moreno}, {Moretti},
  {Neustroev}, {Nigro}, {Nilsson}, {Ninci}, {Nishijima}, {Noda}, {Nozaki},
  {Ohtani}, {Oka}, {Otero-Santos}, {Paiano}, {Palatiello}, {Paneque},
  {Paoletti}, {Paredes}, {Pavleti{\'c}}, {Pe{\~n}il}, {Perennes}, {Persic},
  {Prada Moroni}, {Prandini}, {Priyadarshi}, {Puljak}, {Rhode}, {Rib{\'o}},
  {Rico}, {Righi}, {Rugliancich}, {Saha}, {Sahakyan}, {Saito}, {Sakurai},
  {Satalecka}, {Saturni}, {Schleicher}, {Schmidt}, {Schweizer}, {Sitarek},
  {{\v{S}}nidari{\'c}}, {Sobczynska}, {Spolon}, {Stamerra}, {Strom}, {Strzys},
  {Suda}, {Suri{\'c}}, {Takahashi}, {Tavecchio}, {Temnikov}, {Terzi{\'c}},
  {Teshima}, {Torres-Alb{\`a}}, {Tosti}, {Truzzi}, {Tutone}, {van
  Scherpenberg}, {Vanzo}, {Vazquez Acosta}, {Ventura}, {Verguilov}, {Vigorito},
  {Vitale}, {Vovk}, {Will}, {Zari{\'c}}, {MAGIC Collaboration}, \&
  {Nava}}]{Acciari-2021-Ansoldi-ApJ...908...90A}
{Acciari}, V.~A., {Ansoldi}, S., {Antonelli}, L.~A., {et~al.} 2021, \apj, 908,
  90, \dodoi{10.3847/1538-4357/abd249}

\bibitem[{{Aleksi{\'c}} {et~al.}(2016){Aleksi{\'c}}, {Ansoldi}, {Antonelli},
  {Antoranz}, {Babic}, {Bangale}, {Barcel{\'o}}, {Barrio}, {Becerra
  Gonz{\'a}lez}, {Bednarek}, {Bernardini}, {Biasuzzi}, {Biland}, {Bitossi},
  {Blanch}, {Bonnefoy}, {Bonnoli}, {Borracci}, {Bretz}, {Carmona}, {Carosi},
  {Cecchi}, {Colin}, {Colombo}, {Contreras}, {Corti}, {Cortina}, {Covino}, {Da
  Vela}, {Dazzi}, {De Angelis}, {De Caneva}, {De Lotto}, {de O{\~n}a Wilhelmi},
  {Delgado Mendez}, {Dettlaff}, {Dominis Prester}, {Dorner}, {Doro}, {Einecke},
  {Eisenacher}, {Elsaesser}, {Fidalgo}, {Fink}, {Fonseca}, {Font}, {Frantzen},
  {Fruck}, {Galindo}, {Garc{\'\i}a L{\'o}pez}, {Garczarczyk}, {Garrido
  Terrats}, {Gaug}, {Giavitto}, {Godinovi{\'c}}, {Gonz{\'a}lez Mu{\~n}oz},
  {Gozzini}, {Haberer}, {Hadasch}, {Hanabata}, {Hayashida}, {Herrera},
  {Hildebrand}, {Hose}, {Hrupec}, {Idec}, {Illa}, {Kadenius}, {Kellermann},
  {Knoetig}, {Kodani}, {Konno}, {Krause}, {Kubo}, {Kushida}, {La Barbera},
  {Lelas}, {Lemus}, {Lewandowska}, {Lindfors}, {Lombardi}, {Longo},
  {L{\'o}pez}, {L{\'o}pez-Coto}, {L{\'o}pez-Oramas}, {Lorca}, {Lorenz},
  {Lozano}, {Makariev}, {Mallot}, {Maneva}, {Mankuzhiyil}, {Mannheim},
  {Maraschi}, {Marcote}, {Mariotti}, {Mart{\'\i}nez}, {Mazin}, {Menzel},
  {Miranda}, {Mirzoyan}, {Moralejo}, {Munar-Adrover}, {Nakajima}, {Negrello},
  {Neustroev}, {Niedzwiecki}, {Nilsson}, {Nishijima}, {Noda}, {Orito},
  {Overkemping}, {Paiano}, {Palatiello}, {Paneque}, {Paoletti}, {Paredes},
  {Paredes-Fortuny}, {Persic}, {Poutanen}, {Prada Moroni}, {Prandini},
  {Puljak}, {Reinthal}, {Rhode}, {Rib{\'o}}, {Rico}, {Rodriguez Garcia},
  {R{\"u}gamer}, {Saito}, {Saito}, {Satalecka}, {Scalzotto}, {Scapin},
  {Schultz}, {Schlammer}, {Schmidl}, {Schweizer}, {Shore}, {Sillanp{\"a}{\"a}},
  {Sitarek}, {Snidaric}, {Sobczynska}, {Spanier}, {Stamerra}, {Steinbring},
  {Storz}, {Strzys}, {Takalo}, {Takami}, {Tavecchio}, {Tejedor}, {Temnikov},
  {Terzi{\'c}}, {Tescaro}, {Teshima}, {Thaele}, {Tibolla}, {Torres}, {Toyama},
  {Treves}, {Vogler}, {Wetteskind}, {Will}, \&
  {Zanin}}]{Aleksic_2016_Ansoldi_APh....72...76A}
{Aleksi{\'c}}, J., {Ansoldi}, S., {Antonelli}, L.~A., {et~al.} 2016,
  Astroparticle Physics, 72, 76, \dodoi{10.1016/j.astropartphys.2015.02.005}

\bibitem[{{Ayala}(2020)}]{Ayala-2020-GCN.29086....1A}
{Ayala}, H. 2020, GRB Coordinates Network, 29086, 1

\bibitem[{{Band} {et~al.}(1993){Band}, {Matteson}, {Ford}, {Schaefer},
  {Palmer}, {Teegarden}, {Cline}, {Briggs}, {Paciesas}, {Pendleton}, {Fishman},
  {Kouveliotou}, {Meegan}, {Wilson}, \&
  {Lestrade}}]{Band-1993-Matteson-ApJ...413..281B}
{Band}, D., {Matteson}, J., {Ford}, L., {et~al.} 1993, \apj, 413, 281,
  \dodoi{10.1086/172995}

\bibitem[{Barbary(2016)}]{Barbary_2016___v.p..}
Barbary, K. 2016, Extinction V0.3.0,  Zenodo, \dodoi{10.5281/ZENODO.804966}

\bibitem[{{Beardmore} {et~al.}(2020){Beardmore}, {Gropp}, {Kennea}, {Klingler},
  {Laha}, {Lien}, {Moss}, {Page}, {Palmer}, {Tohuvavohu}, \& {Neil Gehrels
  Swift Observatory Team}}]{Beardmore-2020-Gropp-GCN.29061....1B}
{Beardmore}, A.~P., {Gropp}, J.~D., {Kennea}, J.~A., {et~al.} 2020, GRB
  Coordinates Network, 29061, 1

\bibitem[{{Beniamini} {et~al.}(2015){Beniamini}, {Nava}, {Duran}, \&
  {Piran}}]{Beniamini-2015-Nava-MNRAS.454.1073B}
{Beniamini}, P., {Nava}, L., {Duran}, R.~B., \& {Piran}, T. 2015, \mnras, 454,
  1073, \dodoi{10.1093/mnras/stv2033}

\bibitem[{{Bissaldi} {et~al.}(2020){Bissaldi}, {Omodei}, {Kocevski},
  {Axelsson}, {Longo}, {Moretti}, \& {Fermi-LAT
  Collaboration}}]{Bissaldi-2020-Omodei-GCN.29076....1B}
{Bissaldi}, E., {Omodei}, N., {Kocevski}, D., {et~al.} 2020, GRB Coordinates
  Network, 29076, 1

\bibitem[{{Blanch} {et~al.}(2020{\natexlab{a}}){Blanch}, {Gaug}, {Noda},
  {Berti}, {Moretti}, {Miceli}, {Gliwny}, {Ubach}, {Schleicher}, {Cerruti},
  {Stamerra}, \& {MAGIC Collaboration}}]{Blanch-2020-Gaug-GCN.28659....1B}
{Blanch}, O., {Gaug}, M., {Noda}, K., {et~al.} 2020{\natexlab{a}}, GRB
  Coordinates Network, 28659, 1

\bibitem[{{Blanch} {et~al.}(2020{\natexlab{b}}){Blanch}, {Longo}, {Berti},
  {Fukami}, {Suda}, {Loporchio}, {Micanovic}, {Green}, {Pinter}, {Takahashi},
  \& {MAGIC Collaboration}}]{Blanch-2020-Longo-GCN.29075....1B}
{Blanch}, O., {Longo}, F., {Berti}, A., {et~al.} 2020{\natexlab{b}}, GRB
  Coordinates Network, 29075, 1

\bibitem[{Buchner(2016)}]{Buchner_2016__StatisticsandComputing_v26.p383..392}
Buchner, J. 2016, Statistics and Computing, 26, 383,
  \dodoi{10.1007/s11222-014-9512-y}

\bibitem[{Buchner(2019)}]{Buchner_2019__pasp_v131.p108005..108005}
---. 2019, \pasp, 131, 108005, \dodoi{10.1088/1538-3873/aae7fc}

\bibitem[{Buchner(2021)}]{Buchner_2021__TheJournalofOpenSourceSoftware_v6.p3001..3001}
---. 2021, The Journal of Open Source Software, 6, 3001,
  \dodoi{10.21105/joss.03001}

\bibitem[{{Campana} {et~al.}(2020){Campana}, {Cusumano}, {Evans}, {Osborne},
  {Burrows}, {Kennea}, \& {Swift-XRT
  Team}}]{Campana-2020-Cusumano-GCN.29064....1C}
{Campana}, S., {Cusumano}, G., {Evans}, P.~A., {et~al.} 2020, GRB Coordinates
  Network, 29064, 1

\bibitem[{{D'Ai} {et~al.}(2020){D'Ai}, {Gropp}, {Kennea}, {Tohuvavohu}, {Page},
  {Beardmore}, {Evans}, {Melandri}, {D'Elia}, \& {Swift-XRT
  Team}}]{D'Ai-2020-Gropp-GCN.28660....1D}
{D'Ai}, A., {Gropp}, J.~D., {Kennea}, J.~A., {et~al.} 2020, GRB Coordinates
  Network, 28660, 1

\bibitem[{{D'Elia} \& {Swift
  Team}(2020)}]{D'Elia-2020-Swift_Team-GCN.28857....1D}
{D'Elia}, V., \& {Swift Team}. 2020, GRB Coordinates Network, 28857, 1

\bibitem[{{Dermer} {et~al.}(2000){Dermer}, {Chiang}, \&
  {Mitman}}]{Dermer_2000_Chiang_ApJ...537..785D}
{Dermer}, C.~D., {Chiang}, J., \& {Mitman}, K.~E. 2000, \apj, 537, 785,
  \dodoi{10.1086/309061}

\bibitem[{Dom{\'\i}nguez {et~al.}(2011)Dom{\'\i}nguez, Primack, Rosario, Prada,
  Gilmore, Faber, Koo, Somerville, P{\'e}rez-Torres, P{\'e}rez-Gonz{\'a}lez,
  Huang, Davis, Guhathakurta, Barmby, Conselice, Lozano, Newman, \&
  Cooper}]{Dominguez_2011_Primack_mnras_v410.p2556..2578}
Dom{\'\i}nguez, A., Primack, J.~R., Rosario, D.~J., {et~al.} 2011, \mnras, 410,
  2556, \dodoi{10.1111/j.1365-2966.2010.17631.x}

\bibitem[{{Fan} \& {Piran}(2006)}]{Fan-2006-Piran--MNRAS.369..197F}
{Fan}, Y., \& {Piran}, T. 2006, \mnras, 369, 197,
  \dodoi{10.1111/j.1365-2966.2006.10280.x}

\bibitem[{{Fan} \& {Piran}(2008)}]{Fan_2008_Piran_FrPhC...3..306F}
{Fan}, Y.-Z., \& {Piran}, T. 2008, Frontiers of Physics in China, 3, 306,
  \dodoi{10.1007/s11467-008-0033-z}

\bibitem[{Fan {et~al.}(2008)Fan, Piran, Narayan, \&
  Wei}]{Fan_2008_Piran_mnras_v384.p1483..1501}
Fan, Y.~Z., Piran, T., Narayan, R., \& Wei, D.-M. 2008, \mnras, 384, 1483,
  \dodoi{10.1111/j.1365-2966.2007.12765.x}

\bibitem[{{Fermi GBM Team}(2020)}]{Fermi_GBM_Team_2020GCN.29063....1F}
{Fermi GBM Team}. 2020, GRB Coordinates Network, 29063, 1

\bibitem[{{Fitzpatrick}(1999)}]{Fitzpatrick-1999-PASP..111...63F}
{Fitzpatrick}, E.~L. 1999, \pasp, 111, 63, \dodoi{10.1086/316293}

\bibitem[{{Fletcher} {et~al.}(2020){Fletcher}, {Veres}, \& {Fermi-GBM
  Team}}]{Fletcher-2020-Veres-GCN.28663....1F}
{Fletcher}, C., {Veres}, P., \& {Fermi-GBM Team}. 2020, GRB Coordinates
  Network, 28663, 1

\bibitem[{Foreman-Mackey(2016)}]{ForemanMackey_2016__TheJournalofOpenSourceSoftware_v1.p24..24}
Foreman-Mackey, D. 2016, The Journal of Open Source Software, 1, 24,
  \dodoi{10.21105/joss.00024}

\bibitem[{{Foreman-Mackey} {et~al.}(2013){Foreman-Mackey}, {Hogg}, {Lang}, \&
  {Goodman}}]{Foreman_2013_Hogg-PASP..125..306F}
{Foreman-Mackey}, D., {Hogg}, D.~W., {Lang}, D., \& {Goodman}, J. 2013, \pasp,
  125, 306, \dodoi{10.1086/670067}

\bibitem[{Fukami {et~al.}(2021)Fukami, Berti, Loporchio, Suda, Nava, Noda,
  Bošnjak, Asano, Longo, Acciari, Ansoldi, Antonelli, Arbet~Engels, Artero,
  Baack, Babić, Baquero, Barres~de Almeida, Barrio, Batković,
  Becerra~González, Bednarek, Bellizzi, Bernardini, Bernardos, Besenrieder,
  Bhattacharyya, Bigongiari, Biland, Blanch, Bökenkamp, Bonnoli, Busetto,
  Carosi, Ceribella, Cerruti, Chai, Chilingarian, Cikota, Colak, Colombo,
  Contreras, Cortina, Covino, D'Amico, D'Elia, Da~Vela, Dazzi, De~Angelis,
  De~Lotto, Delfino, Delgado, Delgado~Mendez, Depaoli, Di~Pierro, Di~Venere,
  Do~Souto~Espiñeira, Dominis~Prester, Donini, Dorner, Doro, Elsaesser,
  Fallah~Ramazani, Fattorini, Fonseca, Font, Fruck, Fukazawa, García~López,
  Garczarczyk, Gasparyan, Gaug, Giglietto, Giordano, Gliwny, Godinović, Green,
  Green, Hadasch, Hahn, Heckmann, Herrera, Hoang, Hrupec, Hütten, Inada,
  Ishio, Iwamura, Jiménez~Martínez, Jormanainen, Jouvin, Karjalainen,
  Kerszberg, Kobayashi, Kubo, Kushida, Lamastra, LELAS, Leone, Lindfors,
  Linhoff, Lombardi, López-Coto, López-Moya, López-Oramas, Machado~de
  Oliveira~Fraga, Maggio, Majumdar, MAKARIEV, Mallamaci, Maneva, Manganaro,
  Mannheim, Maraschi, Mariotti, Martínez, Mazin, Menchiari, Mender,
  Mićanović, Miceli, Miener, Miranda, Mirzoyan, Molina, Moralejo, Morcuende,
  Moreno, Moretti, Nakamori, Neustroev, Nigro, Nilsson, Nishijima, Nozaki,
  Ohtani, Oka, Otero-Santos, Paiano, Palatiello, Paneque, Paoletti, Paredes,
  Pavletić, Peñil, Persic, Pihet, Prada~Moroni, Prandini, Priyadarshi,
  Puljak, Rhode, Ribó, Rico, Righi, Rugliancich, Sahakyan, Saito, Sakurai,
  Satalecka, Saturni, Schleicher, Schmidt, Schweizer, Sitarek, Šnidarić,
  Sobczyńska, Spolon, Stamerra, Strišković, Strom, Strzys, Surić,
  Takahashi, Takeishi, Tavecchio, Temnikov, Terzić, Teshima, Tosti, Truzzi,
  Tutone, Ubach, van Scherpenberg, Vanzo, VAZQUEZ~ACOSTA, Ventura, VERGUILOV,
  Vigorito, Vitale, Vovk, Will, Wunderlich, Yamamoto, \&
  Zarić}]{Fukami-2022-Berti-icrc.confE.788F}
Fukami, S., Berti, A., Loporchio, S., {et~al.} 2021, in Proceedings of 37th
  International Cosmic Ray Conference {\textemdash} PoS(ICRC2021), Vol. 395,
  788, \dodoi{10.22323/1.395.0788}

\bibitem[{{Gao} {et~al.}(2015){Gao}, {Wang}, {M{\'e}sz{\'a}ros}, \&
  {Zhang}}]{Gao-2015-Wang-ApJ...810..160G}
{Gao}, H., {Wang}, X.-G., {M{\'e}sz{\'a}ros}, P., \& {Zhang}, B. 2015, \apj,
  810, 160, \dodoi{10.1088/0004-637X/810/2/160}

\bibitem[{Geng {et~al.}(2018)Geng, Huang, Wu, Zhang, \&
  Zong}]{Geng_2018_Huang_apjs_v234.p3..3}
Geng, J.-J., Huang, Y.-F., Wu, X.-F., Zhang, B., \& Zong, H.-S. 2018, \apjs,
  234, 3, \dodoi{10.3847/1538-4365/aa9e84}

\bibitem[{{Giarratana} {et~al.}(2022){Giarratana}, {Rhodes}, {Marcote},
  {Fender}, {Ghirlanda}, {Giroletti}, {Nava}, {Paredes}, {Ravasio}, {Ribo},
  {Patel}, {Rastinejad}, {Schroeder}, {Fong}, {Gompertz}, {Levan}, \&
  {O'Brien}}]{Giarratana-2022-Rhodes-arXiv220512750G}
{Giarratana}, S., {Rhodes}, L., {Marcote}, B., {et~al.} 2022, arXiv e-prints,
  arXiv:2205.12750.
\newblock \doarXiv{2205.12750}

\bibitem[{{Gompertz} {et~al.}(2020){Gompertz}, {Levan}, {Tanvir}, {Fruchter},
  {Cucchiara}, {Greiner}, {Hjorth}, {Kangas}, {Lamb}, {Lyman}, {Oates},
  {O'Brien}, {Osborne}, {Page}, {Perley}, {Pian}, {Steeghs}, {Wiersema}, \&
  {Wynn}}]{Gompertz-2020-Levan-GCN.28822....1G}
{Gompertz}, B., {Levan}, A., {Tanvir}, N., {et~al.} 2020, GRB Coordinates
  Network, 28822, 1

\bibitem[{{Gould} \&
  {Schr{\'e}der}(1967)}]{Gould_1967_Schreder_PhRv..155.1404G}
{Gould}, R.~J., \& {Schr{\'e}der}, G.~P. 1967, Physical Review, 155, 1404,
  \dodoi{10.1103/PhysRev.155.1404}

\bibitem[{Granot \& Sari(2002)}]{Granot_2002_Sari_apj_v568.p820..829}
Granot, J., \& Sari, R. 2002, \apj, 568, 820, \dodoi{10.1086/338966}

\bibitem[{{Greiner} {et~al.}(2015){Greiner}, {Mazzali}, {Kann}, {Kr{\"u}hler},
  {Pian}, {Prentice}, {Olivares E.}, {Rossi}, {Klose}, {Taubenberger}, {Knust},
  {Afonso}, {Ashall}, {Bolmer}, {Delvaux}, {Diehl}, {Elliott}, {Filgas},
  {Fynbo}, {Graham}, {Guelbenzu}, {Kobayashi}, {Leloudas}, {Savaglio},
  {Schady}, {Schmidl}, {Schweyer}, {Sudilovsky}, {Tanga}, {Updike}, {van
  Eerten}, \& {Varela}}]{Greiner_2015_Mazzali_Natur.523..189G}
{Greiner}, J., {Mazzali}, P.~A., {Kann}, D.~A., {et~al.} 2015, \nat, 523, 189,
  \dodoi{10.1038/nature14579}

\bibitem[{{Grossan} {et~al.}(2020){Grossan}, {Maksut}, {Kim}, {Krugov},
  {Smoot}, \& {Linder}}]{Grossan-2020-Maksut-GCN.28674....1G}
{Grossan}, B., {Maksut}, Z., {Kim}, A., {et~al.} 2020, GRB Coordinates Network,
  28674, 1

\bibitem[{{H.~E.~S.~S. Collaboration} {et~al.}(2021){H.~E.~S.~S.
  Collaboration}, {Abdalla}, {Aharonian}, {Ait Benkhali}, {Ang{\"u}ner},
  {Arcaro}, {Armand}, {Armstrong}, {Ashkar}, {Backes}, {Baghmanyan}, {Barbosa
  Martins}, {Barnacka}, {Barnard}, {Becherini}, {Berge}, {Bernl{\"o}hr}, {Bi},
  {Bissaldi}, {B{\"o}ttcher}, {Boisson}, {Bolmont}, {de Bony de Lavergne},
  {Breuhaus}, {Brun}, {Brun}, {Bryan}, {B{\"u}chele}, {Bulik}, {Bylund},
  {Caroff}, {Carosi}, {Casanova}, {Chand}, {Chandra}, {Chen}, {Cotter},
  {Cury{\l}o}, {Damascene Mbarubucyeye}, {Davids}, {Davies}, {Deil}, {Devin},
  {Dirson}, {Djannati-Ata{\"\i}}, {Dmytriiev}, {Donath}, {Doroshenko},
  {Dreyer}, {Duffy}, {Dyks}, {Egberts}, {Eichhorn}, {Einecke}, {Emery},
  {Ernenwein}, {Feijen}, {Fegan}, {Fiasson}, {Fichet de Clairfontaine},
  {Fontaine}, {Funk}, {F{\"u}{\ss}ling}, {Gabici}, {Gallant}, {Giavitto},
  {Giunti}, {Glawion}, {Glicenstein}, {Grondin}, {Hahn}, {Haupt}, {Hermann},
  {Hinton}, {Hofmann}, {Hoischen}, {Holch}, {Holler}, {H{\"o}rbe}, {Horns},
  {Huber}, {Jamrozy}, {Jankowsky}, {Jankowsky}, {Jardin-Blicq}, {Joshi},
  {Jung-Richardt}, {Kasai}, {Kastendieck}, {Katarzy{\'n}ski}, {Katz},
  {Khangulyan}, {Kh{\'e}lifi}, {Klepser}, {Klu{\'z}niak}, {Komin}, {Konno},
  {Kosack}, {Kostunin}, {Kreter}, {Lamanna}, {Lemi{\`e}re}, {Lemoine-Goumard},
  {Lenain}, {Leuschner}, {Levy}, {Lohse}, {Lypova}, {Mackey}, {Majumdar},
  {Malyshev}, {Malyshev}, {Marandon}, {Marchegiani}, {Marcowith}, {Mares},
  {Mart{\'\i}-Devesa}, {Marx}, {Maurin}, {Meintjes}, {Meyer}, {Mitchell},
  {Moderski}, {Mohrmann}, {Montanari}, {Moore}, {Morris}, {Moulin}, {Muller},
  {Murach}, {Nakashima}, {Nayerhoda}, {de Naurois}, {Ndiyavala}, {Niemiec},
  {Oakes}, {O'Brien}, {Odaka}, {Ohm}, {Olivera-Nieto}, {de Ona Wilhelmi},
  {Ostrowski}, {Panny}, {Panter}, {Parsons}, {Peron}, {Peyaud}, {Piel}, {Pita},
  {Poireau}, {Priyana Noel}, {Prokhorov}, {Prokoph}, {P{\"u}hlhofer}, {Punch},
  {Quirrenbach}, {Raab}, {Rauth}, {Reichherzer}, {Reimer}, {Reimer}, {Remy},
  {Renaud}, {Rieger}, {Rinchiuso}, {Romoli}, {Rowell}, {Rudak}, {Ruiz-Velasco},
  {Sahakian}, {Sailer}, {Salzmann}, {Sanchez}, {Santangelo}, {Sasaki},
  {Scalici}, {Sch{\"a}fer}, {Sch{\"u}ssler}, {Schutte}, {Schwanke},
  {Seglar-Arroyo}, {Senniappan}, {Seyffert}, {Shafi}, {Shapopi},
  {Shiningayamwe}, {Simoni}, {Sinha}, {Sol}, {Specovius}, {Spencer},
  {Spir-Jacob}, {Stawarz}, {Sun}, {Steenkamp}, {Stegmann}, {Steinmassl},
  {Steppa}, {Takahashi}, {Tam}, {Tavernier}, {Taylor}, {Terrier}, {Thiersen},
  {Tiziani}, {Tluczykont}, {Tomankova}, {Tsirou}, {Tuffs}, {Uchiyama}, {van der
  Walt}, {van Eldik}, {van Rensburg}, {van Soelen}, {Vasileiadis}, {Veh},
  {Venter}, {Vincent}, {Vink}, {V{\"o}lk}, {Wadiasingh}, {Wagner}, {Watson},
  {Werner}, {White}, {Wierzcholska}, {Wong}, {Yusafzai}, {Zacharias}, {Zanin},
  {Zargaryan}, {Zdziarski}, {Zech}, {Zhu}, {Zorn}, {Zouari}, {{\.Z}ywucka},
  {Evans}, \& {Page}}]{H._E._S._S._Collaboration_2021_Abdalla_Sci...372.1081H}
{H.~E.~S.~S. Collaboration}, {Abdalla}, H., {Aharonian}, F., {et~al.} 2021,
  Science, 372, 1081, \dodoi{10.1126/science.abe8560}

\bibitem[{{Hu} {et~al.}(2021){Hu}, {Castro-Tirado}, {Kumar}, {Gupta}, {Valeev},
  {Pandey}, {Kann}, {Castell{\'o}n}, {Agudo}, {Aryan}, {Caballero-Garc{\'\i}a},
  {Guziy}, {Martin-Carrillo}, {Oates}, {Pian}, {S{\'a}nchez-Ram{\'\i}rez},
  {Sokolov}, \& {Zhang}}]{Hu-2021-Castro-Tirado-A&A...646A..50H}
{Hu}, Y.~D., {Castro-Tirado}, A.~J., {Kumar}, A., {et~al.} 2021, \aap, 646,
  A50, \dodoi{10.1051/0004-6361/202039349}

\bibitem[{{Huang} {et~al.}(2020){Huang}, {Liang}, {Liu}, {Cheng}, \&
  {Wang}}]{Huang_2020_Liang_ApJ...903L..26H}
{Huang}, X.-L., {Liang}, E.-W., {Liu}, R.-Y., {Cheng}, J.-G., \& {Wang}, X.-Y.
  2020, \apjl, 903, L26, \dodoi{10.3847/2041-8213/abc330}

\bibitem[{{Huang} {et~al.}(2000){Huang}, {Gou}, {Dai}, \&
  {Lu}}]{Huang-2000-Gou-ApJ...543...90H}
{Huang}, Y.~F., {Gou}, L.~J., {Dai}, Z.~G., \& {Lu}, T. 2000, \apj, 543, 90,
  \dodoi{10.1086/317076}

\bibitem[{{Huang} {et~al.}(2022){Huang}, {Kirk}, {Giacinti}, \&
  {Reville}}]{Huang_2022_Kirk_ApJ...925..182H}
{Huang}, Z.-Q., {Kirk}, J.~G., {Giacinti}, G., \& {Reville}, B. 2022, \apj,
  925, 182, \dodoi{10.3847/1538-4357/ac3f38}

\bibitem[{{Hurley} {et~al.}(1994){Hurley}, {Dingus}, {Mukherjee}, {Sreekumar},
  {Kouveliotou}, {Meegan}, {Fishman}, {Band}, {Ford}, {Bertsch}, {Cline},
  {Fichtel}, {Hartman}, {Hunter}, {Thompson}, {Kanbach}, {Mayer-Hasselwander},
  {von Montigny}, {Sommer}, {Lin}, {Nolan}, {Michelson}, {Kniffen}, {Mattox},
  {Schneid}, {Boer}, \& {Niel}}]{Hurley_1994_Dingus_Natur.372..652H}
{Hurley}, K., {Dingus}, B.~L., {Mukherjee}, R., {et~al.} 1994, \nat, 372, 652,
  \dodoi{10.1038/372652a0}

\bibitem[{{Izzo} {et~al.}(2020{\natexlab{a}}){Izzo}, {Malesani}, \&
  {Kann}}]{Izzo-2020-Malesani-GCN.29066....1I}
{Izzo}, L., {Malesani}, D.~B., \& {Kann}, D.~A. 2020{\natexlab{a}}, GRB
  Coordinates Network, 29066, 1

\bibitem[{{Izzo} {et~al.}(2020{\natexlab{b}}){Izzo}, {Malesani}, {Zhu}, {Xu},
  {de Ugarte Postigo}, \& {Pursimo}}]{Izzo-2020-Malesani-GCN.28661....1I}
{Izzo}, L., {Malesani}, D.~B., {Zhu}, Z.~P., {et~al.} 2020{\natexlab{b}}, GRB
  Coordinates Network, 28661, 1

\bibitem[{{Joshi} \& {Razzaque}(2021)}]{Joshi_2021_Razzaque_MNRAS.505.1718J}
{Joshi}, J.~C., \& {Razzaque}, S. 2021, \mnras, 505, 1718,
  \dodoi{10.1093/mnras/stab1329}

\bibitem[{{Komesh} {et~al.}(2022){Komesh}, {Grossan}, {Maksut}, {Abdikamalov},
  {Krugov}, \& {Smoot}}]{Komesh-2022-Grossan-arXiv221103029K}
{Komesh}, T., {Grossan}, B., {Maksut}, Z., {et~al.} 2022, arXiv e-prints,
  arXiv:2211.03029.
\newblock \doarXiv{2211.03029}

\bibitem[{Kumar {et~al.}(2012)Kumar, Hern{\'a}ndez, Bo{\v{s}}njak, \&
  Barniol~Duran}]{Kumar_2012_Hernandez_mnras_v427.p40..44L}
Kumar, P., Hern{\'a}ndez, R.~A., Bo{\v{s}}njak, {\v{Z}}., \& Barniol~Duran, R.
  2012, \mnras, 427, L40, \dodoi{10.1111/j.1745-3933.2012.01341.x}

\bibitem[{{Liang} {et~al.}(2007){Liang}, {Zhang}, {Virgili}, \&
  {Dai}}]{Liang_2007_Zhang_ApJ...662.1111L}
{Liang}, E., {Zhang}, B., {Virgili}, F., \& {Dai}, Z.~G. 2007, \apj, 662, 1111,
  \dodoi{10.1086/517959}

\bibitem[{{Liang} {et~al.}(2015){Liang}, {Lin}, {L{\"u}}, {Lu}, {Zhang}, \&
  {Zhang}}]{Liang_2015_Lin_ApJ...813..116L}
{Liang}, E.-W., {Lin}, T.-T., {L{\"u}}, J., {et~al.} 2015, \apj, 813, 116,
  \dodoi{10.1088/0004-637X/813/2/116}

\bibitem[{{Lipunov} {et~al.}(2020){Lipunov}, {Gorbovskoy}, {Kornilov},
  {Tyurina}, {Balanutsa}, {Kuznetsov}, {Balakin}, {Vladimirov}, {Vlasenko},
  {Gorbunov}, {Zimnukhov}, {Senik}, {Pogrosheva}, {Kuvshinov}, {Cheryasov},
  {Podesta}, {Lopez}, {Podesta}, {Francile}, {Rebolo}, {Serra}, {Buckley},
  {Gres}, {Budnev}, {Ershova}, {Tlatov}, {Dormidontov}, {Yurkov}, {Gabovich},
  \& {Sergienko}}]{Lipunov_2020_Gorbovskoy_TNSTR3144....1L}
{Lipunov}, V., {Gorbovskoy}, E., {Kornilov}, V., {et~al.} 2020, Transient Name
  Server Discovery Report, 2020-3144, 1

\bibitem[{{MAGIC Collaboration} {et~al.}(2019{\natexlab{a}}){MAGIC
  Collaboration}, {Acciari}, {Ansoldi}, {Antonelli}, {Arbet Engels}, {Baack},
  {Babi{\'c}}, {Banerjee}, {Barres de Almeida}, {Barrio}, {Becerra
  Gonz{\'a}lez}, {Bednarek}, {Bellizzi}, {Bernardini}, {Berti}, {Besenrieder},
  {Bhattacharyya}, {Bigongiari}, {Biland}, {Blanch}, {Bonnoli},
  {Bo{\v{s}}njak}, {Busetto}, {Carosi}, {Carosi}, {Ceribella}, {Chai},
  {Chilingaryan}, {Cikota}, {Colak}, {Colin}, {Colombo}, {Contreras},
  {Cortina}, {Covino}, {D'Amico}, {D'Elia}, {da Vela}, {Dazzi}, {de Angelis},
  {de Lotto}, {Delfino}, {Delgado}, {Depaoli}, {di Pierro}, {di Venere}, {Do
  Souto Espi{\~n}eira}, {Dominis Prester}, {Donini}, {Dorner}, {Doro},
  {Elsaesser}, {Fallah Ramazani}, {Fattorini}, {Fern{\'a}ndez-Barral},
  {Ferrara}, {Fidalgo}, {Foffano}, {Fonseca}, {Font}, {Fruck}, {Fukami},
  {Gallozzi}, {Garc{\'\i}a L{\'o}pez}, {Garczarczyk}, {Gasparyan}, {Gaug},
  {Giglietto}, {Giordano}, {Godinovi{\'c}}, {Green}, {Guberman}, {Hadasch},
  {Hahn}, {Herrera}, {Hoang}, {Hrupec}, {H{\"u}tten}, {Inada}, {Inoue},
  {Ishio}, {Iwamura}, {Jouvin}, {Kerszberg}, {Kubo}, {Kushida}, {Lamastra},
  {Lelas}, {Leone}, {Lindfors}, {Lombardi}, {Longo}, {L{\'o}pez},
  {L{\'o}pez-Coto}, {L{\'o}pez-Oramas}, {Loporchio}, {Machado de Oliveira
  Fraga}, {Maggio}, {Majumdar}, {Makariev}, {Mallamaci}, {Maneva}, {Manganaro},
  {Mannheim}, {Maraschi}, {Mariotti}, {Mart{\'\i}nez}, {Masuda}, {Mazin},
  {Mi{\'c}anovi{\'c}}, {Miceli}, {Minev}, {Miranda}, {Mirzoyan}, {Molina},
  {Moralejo}, {Morcuende}, {Moreno}, {Moretti}, {Munar-Adrover}, {Neustroev},
  {Nigro}, {Nilsson}, {Ninci}, {Nishijima}, {Noda}, {Nogu{\'e}s}, {N{\"o}the},
  {Nozaki}, {Paiano}, {Palacio}, {Palatiello}, {Paneque}, {Paoletti},
  {Paredes}, {Pe{\~n}il}, {Peresano}, {Persic}, {Prada Moroni}, {Prandini},
  {Puljak}, {Rhode}, {Rib{\'o}}, {Rico}, {Righi}, {Rugliancich}, {Saha},
  {Sahakyan}, {Saito}, {Sakurai}, {Satalecka}, {Schmidt}, {Schweizer},
  {Sitarek}, {{\v{S}}nidari{\'c}}, {Sobczynska}, {Somero}, {Stamerra}, {Strom},
  {Strzys}, {Suda}, {Suri{\'c}}, {Takahashi}, {Tavecchio}, {Temnikov},
  {Terzi{\'c}}, {Teshima}, {Torres-Alb{\`a}}, {Tosti}, {Tsujimoto}, {Vagelli},
  {van Scherpenberg}, {Vanzo}, {Vazquez Acosta}, {Vigorito}, {Vitale}, {Vovk},
  {Will}, {Zari{\'c}}, \&
  {Nava}}]{MAGIC_Collaboration_2019_Acciari_Natur.575..455M}
{MAGIC Collaboration}, {Acciari}, V.~A., {Ansoldi}, S., {et~al.}
  2019{\natexlab{a}}, \nat, 575, 455, \dodoi{10.1038/s41586-019-1750-x}

\bibitem[{{MAGIC Collaboration} {et~al.}(2019{\natexlab{b}}){MAGIC
  Collaboration}, {Acciari}, {Ansoldi}, {Antonelli}, {Engels}, {Baack},
  {Babi{\'c}}, {Banerjee}, {Barres de Almeida}, {Barrio}, {Becerra
  Gonz{\'a}lez}, {Bednarek}, {Bellizzi}, {Bernardini}, {Berti}, {Besenrieder},
  {Bhattacharyya}, {Bigongiari}, {Biland}, {Blanch}, {Bonnoli},
  {Bo{\v{s}}njak}, {Busetto}, {Carosi}, {Ceribella}, {Chai}, {Chilingaryan},
  {Cikota}, {Colak}, {Colin}, {Colombo}, {Contreras}, {Cortina}, {Covino},
  {D'Elia}, {da Vela}, {Dazzi}, {de Angelis}, {de Lotto}, {Delfino}, {Delgado},
  {Depaoli}, {di Pierro}, {di Venere}, {Do Souto Espi{\~n}eira}, {Dominis
  Prester}, {Donini}, {Dorner}, {Doro}, {Elsaesser}, {Fallah Ramazani},
  {Fattorini}, {Ferrara}, {Fidalgo}, {Foffano}, {Fonseca}, {Font}, {Fruck},
  {Fukami}, {Garc{\'\i}a L{\'o}pez}, {Garczarczyk}, {Gasparyan}, {Gaug},
  {Giglietto}, {Giordano}, {Godinovi{\'c}}, {Green}, {Guberman}, {Hadasch},
  {Hahn}, {Herrera}, {Hoang}, {Hrupec}, {H{\"u}tten}, {Inada}, {Inoue},
  {Ishio}, {Iwamura}, {Jouvin}, {Kerszberg}, {Kubo}, {Kushida}, {Lamastra},
  {Lelas}, {Leone}, {Lindfors}, {Lombardi}, {Longo}, {L{\'o}pez},
  {L{\'o}pez-Coto}, {L{\'o}pez-Oramas}, {Loporchio}, {Machado de Oliveira
  Fraga}, {Maggio}, {Majumdar}, {Makariev}, {Mallamaci}, {Maneva}, {Manganaro},
  {Mannheim}, {Maraschi}, {Mariotti}, {Mart{\'\i}nez}, {Mazin},
  {Mi{\'c}anovi{\'c}}, {Miceli}, {Minev}, {Miranda}, {Mirzoyan}, {Molina},
  {Moralejo}, {Morcuende}, {Moreno}, {Moretti}, {Munar-Adrover}, {Neustroev},
  {Nigro}, {Nilsson}, {Ninci}, {Nishijima}, {Noda}, {Nogu{\'e}s}, {Nozaki},
  {Paiano}, {Palatiello}, {Paneque}, {Paoletti}, {Paredes}, {Pe{\~n}il},
  {Peresano}, {Persic}, {Moroni}, {Prandini}, {Puljak}, {Rhode}, {Rib{\'o}},
  {Rico}, {Righi}, {Rugliancich}, {Saha}, {Sahakyan}, {Saito}, {Sakurai},
  {Satalecka}, {Schmidt}, {Schweizer}, {Sitarek}, {{\v{S}}nidari{\'c}},
  {Sobczynska}, {Somero}, {Stamerra}, {Strom}, {Strzys}, {Suda}, {Suri{\'c}},
  {Takahashi}, {Tavecchio}, {Temnikov}, {Terzi{\'c}}, {Teshima},
  {Torres-Alb{\`a}}, {Tosti}, {Vagelli}, {van Scherpenberg}, {Vanzo}, {Vazquez
  Acosta}, {Vigorito}, {Vitale}, {Vovk}, {Will}, {Zari{\'c}}, {Nava}, {Veres},
  {Bhat}, {Briggs}, {Cleveland}, {Hamburg}, {Hui}, {Mailyan}, {Preece},
  {Roberts}, {von Kienlin}, {Wilson-Hodge}, {Kocevski}, {Arimoto}, {Tak},
  {Asano}, {Axelsson}, {Barbiellini}, {Bissaldi}, {Dirirsa}, {Gill}, {Granot},
  {McEnery}, {Omodei}, {Razzaque}, {Piron}, {Racusin}, {Thompson}, {Campana},
  {Bernardini}, {Kuin}, {Siegel}, {Cenko}, {O'Brien}, {Capalbi}, {Da{\i}}, {de
  Pasquale}, {Gropp}, {Klingler}, {Osborne}, {Perri}, {Starling},
  {Tagliaferri}, {Tohuvavohu}, {Ursi}, {Tavani}, {Cardillo}, {Casentini},
  {Piano}, {Evangelista}, {Verrecchia}, {Pittori}, {Lucarelli}, {Bulgarelli},
  {Parmiggiani}, {Anderson}, {Anderson}, {Bernardi}, {Bolmer},
  {Caballero-Garc{\'\i}a}, {Carrasco}, {Castell{\'o}n}, {Castro Segura},
  {Castro-Tirado}, {Cherukuri}, {Cockeram}, {D'Avanzo}, {di Dato}, {Diretse},
  {Fender}, {Fern{\'a}ndez-Garc{\'\i}a}, {Fynbo}, {Fruchter}, {Greiner},
  {Gromadzki}, {Heintz}, {Heywood}, {van der Horst}, {Hu}, {Inserra}, {Izzo},
  {Jaiswal}, {Jakobsson}, {Japelj}, {Kankare}, {Kann}, {Kouveliotou}, {Klose},
  {Levan}, {Li}, {Lotti}, {Maguire}, {Malesani}, {Manulis}, {Marongiu},
  {Martin}, {Melandri}, {Micha{\l}owski}, {Miller-Jones}, {Misra}, {Moin},
  {Mooley}, {Nasri}, {Nicholl}, {Noschese}, {Novara}, {Pandey}, {Peretti},
  {P{\'e}rez Del Pulgar}, {P{\'e}rez-Torres}, {Perley}, {Piro}, {Ragosta},
  {Resmi}, {Ricci}, {Rossi}, {S{\'a}nchez-Ram{\'\i}rez}, {Selsing}, {Schulze},
  {Smartt}, {Smith}, {Sokolov}, {Stevens}, {Tanvir}, {Th{\"o}ne}, {Tiengo},
  {Tremou}, {Troja}, {de Ugarte Postigo}, {Valeev}, {Vergani}, {Wieringa},
  {Woudt}, {Xu}, {Yaron}, \&
  {Young}}]{MAGIC_Collaboration_2019_Acciari_Natur.575..459M}
---. 2019{\natexlab{b}}, \nat, 575, 459, \dodoi{10.1038/s41586-019-1754-6}

\bibitem[{{Markwardt} {et~al.}(2020){Markwardt}, {Barthelmy}, {Cummings},
  {D'Elia}, {Krimm}, {Laha}, {Lien}, {Palmer}, {Sakamoto}, {Stamatikos}, \&
  {Ukwatta}}]{Markwardt-2020-Barthelmy-GCN.28658....1M}
{Markwardt}, C.~B., {Barthelmy}, S.~D., {Cummings}, J.~R., {et~al.} 2020, GRB
  Coordinates Network, 28658, 1

\bibitem[{{Melandri} {et~al.}(2022){Melandri}, {Izzo}, {Pian}, {Malesani},
  {Della Valle}, {Rossi}, {D'Avanzo}, {Guetta}, {Mazzali}, {Benetti},
  {Masetti}, {Palazzi}, {Savaglio}, {Amati}, {Antonelli}, {Ashall},
  {Bernardini}, {Campana}, {Carini}, {Covino}, {D'Elia}, {de Ugarte Postigo},
  {De Pasquale}, {Filippenko}, {Fruchter}, {Fynbo}, {Giunta}, {Hartmann},
  {Jakobsson}, {Japelj}, {Jonker}, {Kann}, {Lamb}, {Levan}, {Martin-Carrillo},
  {M{\o}ller}, {Piranomonte}, {Pugliese}, {Salvaterra}, {Schulze}, {Starling},
  {Stella}, {Tagliaferri}, {Tanvir}, \&
  {Watson}}]{Melandri-2022-Izzo-A&A...659A..39M}
{Melandri}, A., {Izzo}, L., {Pian}, E., {et~al.} 2022, \aap, 659, A39,
  \dodoi{10.1051/0004-6361/202141788}

\bibitem[{{M{\'e}sz{\'a}ros} {et~al.}(2004){M{\'e}sz{\'a}ros}, {Razzaque}, \&
  {Zhang}}]{Meszaros_2004_Razzaque_NewAR..48..445M}
{M{\'e}sz{\'a}ros}, P., {Razzaque}, S., \& {Zhang}, B. 2004, \nar, 48, 445,
  \dodoi{10.1016/j.newar.2003.12.022}

\bibitem[{{M{\'e}sz{\'a}ros} \&
  {Rees}(1993)}]{Meszaros_1993_Rees_ApJ...405..278M}
{M{\'e}sz{\'a}ros}, P., \& {Rees}, M.~J. 1993, \apj, 405, 278,
  \dodoi{10.1086/172360}

\bibitem[{M{\'e}sz{\'a}ros \&
  Rees(1997)}]{Meszaros_1997_Rees_apj_v476.p232..237}
M{\'e}sz{\'a}ros, P., \& Rees, M.~J. 1997, \apj, 476, 232,
  \dodoi{10.1086/303625}

\bibitem[{{Minaev} \&
  {Pozanenko}(2020)}]{Minaev-2020-Pozanenko-GCN.28668....1M}
{Minaev}, P., \& {Pozanenko}, A. 2020, GRB Coordinates Network, 28668, 1

\bibitem[{{Murase} {et~al.}(2006){Murase}, {Ioka}, {Nagataki}, \&
  {Nakamura}}]{Murase_2006_Ioka_ApJ...651L...5M}
{Murase}, K., {Ioka}, K., {Nagataki}, S., \& {Nakamura}, T. 2006, \apjl, 651,
  L5, \dodoi{10.1086/509323}

\bibitem[{Murase {et~al.}(2011)Murase, Toma, Yamazaki, \&
  M{\'{e}}sz{\'{a}}ros}]{Murase_2011_Toma_apj_v732.p77..77}
Murase, K., Toma, K., Yamazaki, R., \& M{\'{e}}sz{\'{a}}ros, P. 2011, \apj,
  732, 77, \dodoi{10.1088/0004-637x/732/2/77}

\bibitem[{{Nakar} {et~al.}(2009){Nakar}, {Ando}, \&
  {Sari}}]{Nakar_2009_Ando_ApJ...703..675N}
{Nakar}, E., {Ando}, S., \& {Sari}, R. 2009, \apj, 703, 675,
  \dodoi{10.1088/0004-637X/703/1/675}

\bibitem[{{O'Connor} {et~al.}(2023){O'Connor}, {Troja}, {Ryan}, {Beniamini},
  {van Eerten}, {Granot}, {Dichiara}, {Ricci}, {Lipunov}, {Gillanders}, {Gill},
  {Moss}, {Anand}, {Andreoni}, {Becerra}, {Buckley}, {Butler}, {Cenko},
  {Chasovnikov}, {Durbak}, {Francile}, {Hammerstein}, {van der Horst},
  {Kasliwal}, {Kouveliotou}, {Kutyrev}, {Lee}, {Srinivasaragavan}, {Topolev},
  {Watson}, {Yang}, \& {Zhirkov}}]{OConnor_2023_Troja_arXiv230207906O}
{O'Connor}, B., {Troja}, E., {Ryan}, G., {et~al.} 2023, arXiv e-prints,
  arXiv:2302.07906, \dodoi{10.48550/arXiv.2302.07906}

\bibitem[{{Pei}(1992)}]{Pei-1992-ApJ...395..130P}
{Pei}, Y.~C. 1992, \apj, 395, 130, \dodoi{10.1086/171637}

\bibitem[{{Planck Collaboration} {et~al.}(2016){Planck Collaboration}, {Ade},
  {Aghanim}, {Arnaud}, {Ashdown}, {Aumont}, {Baccigalupi}, {Banday},
  {Barreiro}, {Bartlett}, \&
  et~al.}]{PlanckCollaboration_2016_Ade_aap_v594.p13..13A}
{Planck Collaboration}, {Ade}, P.~A.~R., {Aghanim}, N., {et~al.} 2016, \aap,
  594, A13, \dodoi{10.1051/0004-6361/201525830}

\bibitem[{{Pozanenko} {et~al.}(2020){Pozanenko}, {Belkin}, {Volnova},
  {Moskvitin}, {Burhonov}, {Kim}, {Krugov}, {Rumyantsev}, {Klunko},
  {Inasaridze}, {Reva}, {Minaev}, {Pankov}, {Ehgamberdiev}, \& {GRB IKI
  FuN}}]{Pozanenko_2020_Belkin_GCN.29033....1P}
{Pozanenko}, A., {Belkin}, S., {Volnova}, A., {et~al.} 2020, GRB Coordinates
  Network, 29033, 1

\bibitem[{{Racusin} {et~al.}(2008){Racusin}, {Karpov}, {Sokolowski}, {Granot},
  {Wu}, {Pal'Shin}, {Covino}, {van der Horst}, {Oates}, {Schady}, {Smith},
  {Cummings}, {Starling}, {Piotrowski}, {Zhang}, {Evans}, {Holland}, {Malek},
  {Page}, {Vetere}, {Margutti}, {Guidorzi}, {Kamble}, {Curran}, {Beardmore},
  {Kouveliotou}, {Mankiewicz}, {Melandri}, {O'Brien}, {Page}, {Piran},
  {Tanvir}, {Wrochna}, {Aptekar}, {Barthelmy}, {Bartolini}, {Beskin}, {Bondar},
  {Bremer}, {Campana}, {Castro-Tirado}, {Cucchiara}, {Cwiok}, {D'Avanzo},
  {D'Elia}, {Della Valle}, {de Ugarte Postigo}, {Dominik}, {Falcone}, {Fiore},
  {Fox}, {Frederiks}, {Fruchter}, {Fugazza}, {Garrett}, {Gehrels},
  {Golenetskii}, {Gomboc}, {Gorosabel}, {Greco}, {Guarnieri}, {Immler},
  {Jelinek}, {Kasprowicz}, {La Parola}, {Levan}, {Mangano}, {Mazets},
  {Molinari}, {Moretti}, {Nawrocki}, {Oleynik}, {Osborne}, {Pagani}, {Pandey},
  {Paragi}, {Perri}, {Piccioni}, {Ramirez-Ruiz}, {Roming}, {Steele}, {Strom},
  {Testa}, {Tosti}, {Ulanov}, {Wiersema}, {Wijers}, {Winters}, {Zarnecki},
  {Zerbi}, {M{\'e}sz{\'a}ros}, {Chincarini}, \&
  {Burrows}}]{Racusin_2008_Karpov_Natur.455..183R}
{Racusin}, J.~L., {Karpov}, S.~V., {Sokolowski}, M., {et~al.} 2008, \nat, 455,
  183, \dodoi{10.1038/nature07270}

\bibitem[{{Razzaque} {et~al.}(2009){Razzaque}, {Dermer}, \&
  {Finke}}]{Razzaque_2009_Dermer_ApJ...697..483R}
{Razzaque}, S., {Dermer}, C.~D., \& {Finke}, J.~D. 2009, \apj, 697, 483,
  \dodoi{10.1088/0004-637X/697/1/483}

\bibitem[{{Ren} {et~al.}(2020){Ren}, {Lin}, {Zhang}, {Wang}, {Li}, {Wang}, \&
  {Liang}}]{Ren-2020-Lin-ApJ...901L..26R}
{Ren}, J., {Lin}, D.-B., {Zhang}, L.-L., {et~al.} 2020, \apjl, 901, L26,
  \dodoi{10.3847/2041-8213/abb672}

\bibitem[{{Ren} {et~al.}(2022){Ren}, {Wang}, {Zhang}, \&
  {Dai}}]{Ren_2022_Wang_arXiv221010673R}
{Ren}, J., {Wang}, Y., {Zhang}, L.-L., \& {Dai}, Z.-G. 2022, arXiv e-prints,
  arXiv:2210.10673, \dodoi{10.48550/arXiv.2210.10673}

\bibitem[{{Rhodes} {et~al.}(2022){Rhodes}, {van der Horst}, {Fender},
  {Aguilera-Dena}, {Bright}, {Vergani}, \&
  {Williams}}]{Rhodes-2022-van_der_Horst-MNRAS.513.1895R}
{Rhodes}, L., {van der Horst}, A.~J., {Fender}, R., {et~al.} 2022, \mnras, 513,
  1895, \dodoi{10.1093/mnras/stac1057}

\bibitem[{{Ror} {et~al.}(2022){Ror}, {Gupta}, {Jel{\'\i}nek}, {Bhushan Pandey},
  {Castro-Tirado}, {Hu}, {Male{\v{n}}{\'a}kov{\'a}}, {{\v{S}}trobl},
  {Th{\"o}ne}, {Hudec}, {Karpov}, {Kumar}, {Aryan}, {Oates},
  {Fern{\'a}ndez-Garc{\'\i}a}, {P{\'e}rez del Pulgar}, {Caballero-Garc{\'\i}a},
  {Castell{\'o}n}, {Carrasco-Garc{\'\i}a}, {P{\'e}rez-Garc{\'\i}a}, {Reina
  Terol}, \& {Rendon}}]{Ror_2022_Gupta_arXiv221110036R}
{Ror}, A.~K., {Gupta}, R., {Jel{\'\i}nek}, M., {et~al.} 2022, arXiv e-prints,
  arXiv:2211.10036.
\newblock \doarXiv{2211.10036}

\bibitem[{{Sahu} {et~al.}(2022){Sahu}, {Valadez Polanco}, \&
  {Rajpoot}}]{Sahu_2022_Valadez_Polanco_ApJ...929...70S}
{Sahu}, S., {Valadez Polanco}, I.~A., \& {Rajpoot}, S. 2022, \apj, 929, 70,
  \dodoi{10.3847/1538-4357/ac5cc6}

\bibitem[{{Sari} \& {Esin}(2001)}]{Sari_2001_Esin_ApJ...548..787S}
{Sari}, R., \& {Esin}, A.~A. 2001, \apj, 548, 787, \dodoi{10.1086/319003}

\bibitem[{Sari \& Esin(2001)}]{Sari_2001_Esin_apj_v548.p787..799}
Sari, R., \& Esin, A.~A. 2001, \apj, 548, 787, \dodoi{10.1086/319003}

\bibitem[{{Sari} {et~al.}(1998){Sari}, {Piran}, \&
  {Narayan}}]{Sari-1998-Piran-ApJ...497L..17S}
{Sari}, R., {Piran}, T., \& {Narayan}, R. 1998, \apjl, 497, L17,
  \dodoi{10.1086/311269}

\bibitem[{{Scargle} {et~al.}(2013){Scargle}, {Norris}, {Jackson}, \&
  {Chiang}}]{Scargle-2013-Norris-ApJ...764..167S}
{Scargle}, J.~D., {Norris}, J.~P., {Jackson}, B., \& {Chiang}, J. 2013, \apj,
  764, 167, \dodoi{10.1088/0004-637X/764/2/167}

\bibitem[{{Schlafly} \&
  {Finkbeiner}(2011)}]{Schlafly-2011-Finkbeiner-ApJ...737..103S}
{Schlafly}, E.~F., \& {Finkbeiner}, D.~P. 2011, \apj, 737, 103,
  \dodoi{10.1088/0004-637X/737/2/103}

\bibitem[{{Shrestha} {et~al.}(2020){Shrestha}, {Melandri}, {Smith}, {Steele},
  {Kobayashi}, {Mundell}, {Gomboc}, \&
  {Guidorzi}}]{Shrestha-2020-Melandri-GCN.29085....1S}
{Shrestha}, M., {Melandri}, A., {Smith}, R., {et~al.} 2020, GRB Coordinates
  Network, 29085, 1

\bibitem[{Speagle(2020)}]{Speagle_2020__mnras_v493.p3132..3158}
Speagle, J.~S. 2020, \mnras, 493, 3132, \dodoi{10.1093/mnras/staa278}

\bibitem[{Suda {et~al.}(2021)Suda, Artero, Asano, Berti, Nava, Noda, Terauchi,
  Acciari, Ansoldi, Antonelli, Sobczyńska, Spolon, Stamerra, Strišković, Strom, Strzys, Surić,
  Takahashi, Takeishi, Tavecchio, Temnikov, Terzic, Teshima, Tosti, Truzzi,
  Tutone, Ubach, van Scherpenberg, Vanzo, VAZQUEZ~ACOSTA, Ventura, VERGUILOV,
  Vigorito, Vitale, Vovk, Will, Wunderlich, Yamamoto, \&
  Zarić}]{Suda-2022-Artero-icrc.confE.797S}
Suda, Y., Artero, M., Asano, K., {et~al.} 2021, in Proceedings of 37th
  International Cosmic Ray Conference {\textemdash} PoS(ICRC2021), Vol. 395,
  797, \dodoi{10.22323/1.395.0797}

\bibitem[{{Vielfaure} {et~al.}(2020){Vielfaure}, {Izzo}, {Xu}, {Vergani},
  {Malesani}, {de Ugarte Postigo}, {D'Elia}, {Fynbo}, {Kann}, {Levan},
  {Pugliese}, {Tanvir}, {Burgarella}, {Rossi}, \& {Stargate
  Consortium}}]{Vielfaure-2020-Izzo-GCN.29077....1V}
{Vielfaure}, J.~B., {Izzo}, L., {Xu}, D., {et~al.} 2020, GRB Coordinates
  Network, 29077, 1

\bibitem[{{Vreeswijk} {et~al.}(2018){Vreeswijk}, {Kann}, {Heintz}, {de Ugarte
  Postigo}, {Milvang-Jensen}, {Malesani}, {Covino}, {Levan}, \&
  {Pugliese}}]{Vreeswijk_2018_Kann_GCN.22996....1V}
{Vreeswijk}, P.~M., {Kann}, D.~A., {Heintz}, K.~E., {et~al.} 2018, GRB
  Coordinates Network, 22996, 1

\bibitem[{{Wang} {et~al.}(2001){Wang}, {Dai}, \&
  {Lu}}]{Wang_2001_Lu_ApJ...556.1010W}
{Wang}, X.~Y., {Dai}, Z.~G., \& {Lu}, T. 2001, \apj, 556, 1010,
  \dodoi{10.1086/321608}

\bibitem[{{Wang} {et~al.}(2019){Wang}, {Liu}, {Zhang}, {Xi}, \&
  {Zhang}}]{Wang_2019_Liu_ApJ...884..117W}
{Wang}, X.-Y., {Liu}, R.-Y., {Zhang}, H.-M., {Xi}, S.-Q., \& {Zhang}, B. 2019,
  \apj, 884, 117, \dodoi{10.3847/1538-4357/ab426c}

\bibitem[{Wang {et~al.}(2022)Wang, Jiang, \&
  Ren}]{Wang_2022_Jiang_arXiveprints_v.p2205..2982arXiv}
Wang, Y., Jiang, L.-Y., \& Ren, J. 2022, arXiv e-prints, arXiv:2205.02982.
\newblock \doarXiv{2205.02982}

\bibitem[{Warren {et~al.}(2022)Warren, Dainotti, Barkov, Ahlgren, Ito, \&
  Nagataki}]{Warren_2022_Dainotti_apj_v924.p40..40}
Warren, D.~C., Dainotti, M., Barkov, M.~V., {et~al.} 2022, \apj, 924, 40,
  \dodoi{10.3847/1538-4357/ac2f43}

\bibitem[{{Waxman}(1997)}]{Waxman_1997_ApJ...485L...5W}
{Waxman}, E. 1997, \apjl, 485, L5, \dodoi{10.1086/310809}

\bibitem[{Waxman(1997)}]{Waxman_1997__apj_v491.p19..22L}
Waxman, E. 1997, \apj, 491, L19, \dodoi{10.1086/311057}

\bibitem[{{Xin} {et~al.}(2011){Xin}, {Liang}, {Wei}, {Zhang}, {Lv}, {Zheng},
  {Urata}, {Im}, {Wang}, {Qiu}, {Deng}, {Huang}, {Hu}, {Jeon}, {Li}, \&
  {Han}}]{Xin-2011-Liang-MNRAS.410...27X}
{Xin}, L.-P., {Liang}, E.-W., {Wei}, J.-Y., {et~al.} 2011, \mnras, 410, 27,
  \dodoi{10.1111/j.1365-2966.2010.17419.x}

\bibitem[{{Yamasaki} \& {Piran}(2022)}]{Yamasaki_2022_Piran_MNRAS.512.2142Y}
{Yamasaki}, S., \& {Piran}, T. 2022, \mnras, 512, 2142,
  \dodoi{10.1093/mnras/stac483}

\bibitem[{{Yost} {et~al.}(2003){Yost}, {Harrison}, {Sari}, \&
  {Frail}}]{Yost_2003_Harrison_ApJ...597..459Y}
{Yost}, S.~A., {Harrison}, F.~A., {Sari}, R., \& {Frail}, D.~A. 2003, \apj,
  597, 459, \dodoi{10.1086/378288}

\bibitem[{{Zhang} {et~al.}(2006){Zhang}, {Fan}, {Dyks}, {Kobayashi},
  {M{\'e}sz{\'a}ros}, {Burrows}, {Nousek}, \&
  {Gehrels}}]{Zhang_2006_Fan_ApJ...642..354Z}
{Zhang}, B., {Fan}, Y.~Z., {Dyks}, J., {et~al.} 2006, \apj, 642, 354,
  \dodoi{10.1086/500723}

\bibitem[{{Zhang} \&
  {M{\'e}sz{\'a}ros}(2001)}]{Zhang_2001_Meszaros_ApJ...559..110Z}
{Zhang}, B., \& {M{\'e}sz{\'a}ros}, P. 2001, \apj, 559, 110,
  \dodoi{10.1086/322400}

\bibitem[{{Zhang} {et~al.}(2009){Zhang}, {Zhang}, {Virgili}, {Liang}, {Kann},
  {Wu}, {Proga}, {Lv}, {Toma}, {M{\'e}sz{\'a}ros}, {Burrows}, {Roming}, \&
  {Gehrels}}]{Zhang-2009-Zhang-ApJ...703.1696Z}
{Zhang}, B., {Zhang}, B.-B., {Virgili}, F.~J., {et~al.} 2009, \apj, 703, 1696,
  \dodoi{10.1088/0004-637X/703/2/1696}

\bibitem[{{Zhang} {et~al.}(2021{\natexlab{a}}){Zhang}, {Murase}, {Veres}, \&
  {M{\'e}sz{\'a}ros}}]{Zhang_2021_Murase_ApJ...920...55Z}
{Zhang}, B.~T., {Murase}, K., {Veres}, P., \& {M{\'e}sz{\'a}ros}, P.
  2021{\natexlab{a}}, \apj, 920, 55, \dodoi{10.3847/1538-4357/ac0cfc}

\bibitem[{Zhang {et~al.}(2020)Zhang, Christie, Petropoulou, Rueda-Becerril, \&
  Giannios}]{Zhang_2020_Christie_mnras_v496.p974..986}
Zhang, H., Christie, I.~M., Petropoulou, M., Rueda-Becerril, J.~M., \&
  Giannios, D. 2020, \mnras, 496, 974, \dodoi{10.1093/mnras/staa1583}

\bibitem[{{Zhang} {et~al.}(2021{\natexlab{b}}){Zhang}, {Ren}, {Huang}, {Liang},
  {Lin}, \& {Liang}}]{Zhang-2021-Ren-ApJ...917...95Z}
{Zhang}, L.-L., {Ren}, J., {Huang}, X.-L., {et~al.} 2021{\natexlab{b}}, \apj,
  917, 95, \dodoi{10.3847/1538-4357/ac0c7f}

\end{thebibliography}

\newpage

\begin{table}[tbph]
\setlength\tabcolsep{2pt}
\caption{Fitting Results of FS Model about GRBs 201015A and 201216C}
\centering
\begin{tabular}{c|cccccccc}
\hline \hline
GRB & ${\rm log_{10}}E_{\rm k, iso} ({\rm erg}$)  & ${\rm log_{10}}\Gamma_{0}$ & ${\rm log_{10}}n_0 (\rm cm^{-3}$) & $p$ & ${\rm log_{10}}\epsilon_{e}$ & ${\rm log_{10}}\epsilon_{B}$ & $E_{\rm B-V}$ &  ${\rm log_{10}}\theta_{j}$\\
\hline
201015A & $52.21^{+0.15}_{-0.16}$  & $ 1.64^{+0.08}_{-0.06}$ & $3.08^{+0.44}_{-0.52}$ & $2.08^{+0.02}_{-0.02}$  & $-0.15^{+0.07}_{-0.07}$ & $-5.44^{+0.79}_{-0.67}$ & $-$  &  $-0.01$ (fixed)\\
201216C & $55.05^{+0.10}_{-0.07}$ & $2.52^{+0.04}_{-0.03}$  &  $0.68^{+0.15}_{-0.12}$  &   $2.05$ (fixed)  &  $-0.85^{+0.09}_{-0.12}$  & $-6.66^{+0.12}_{-0.14}$ &  $1.26$ (fixed) &  $-0.91^{+0.03}_{-0.03}$\\

\hline \hline
\end{tabular}
\label{tab1}
\end{table}

\newpage

\clearpage

\end{document}